\documentclass[12pt]{article}

\usepackage[top=25truemm,bottom=25truemm,left=20truemm,right=20truemm]{geometry}
\usepackage{amsmath,amssymb}
\usepackage{bm,braket,ascmac}
\usepackage{graphicx}
\usepackage{cite}
\usepackage{authblk}
\usepackage{comment}

\title{Dynamics of Open Tight-binding Model}
\author[1]{Takahisa Fukadai}
\author[1]{Tomohiro Sasamoto}
\affil[1]{{\it Department of Physics, Tokyo Institute of Technology, 2-12-1 Ookayama, Meguro-ku, Tokyo 152-8550, Japan}}

\begin{document}
\maketitle

\abstract{In this paper, we investigate the open tight-binding
model with $N$ sites coupled to two reservoirs on its edges with the
nonequilibrium Green function method to understand effects of open
boundaries. As a result, we obtain an analytical expression of the electron
density in the steady state and in the transient regime for $N
\rightarrow \infty$ and at absolute zero temperature. From the
expression of the electron density in the steady state, we show that the
open boundaries do not affect the qualitative behavior of the electron
density and the phase transition which has been
observed for the isolated tight-binding model also exists in our open
case. Using the expression of the time-dependent electron density, we
show that a two-step decay appears in the relaxation process purely caused by
open boundaries. In addition, we show that the
dependence of speed of convergence on boundary coupling strength
qualitatively changes at a certain value of the boundary
parameters. This is a result of special eigenvalues of
a non-hermitian matrix whose imaginary parts do not vanish even in the
limit $N \rightarrow \infty$.
}

\section{Introduction}
With recent remarkable developments of creating and manipulating experimental
systems such as cold atomic system\cite{bloch2008many,cazalilla2011one}, open quantum
systems have been investigated extensively. Especially, low-dimensional
open quantum system is a field of interest, where we have found interesting phenomena such as nonequilibrium phase transitions by dissipation\cite{morrison2008dynamical,prosen2008quantum,gutierrez2017experimental,fink2018signatures} and topological phase
transitions\cite{rudner2009topological,zeuner2015observation}.  

To investigate low-dimensional open quantum systems,
two approaches have been mainly used. One approach is known as Quantum
Master Equation(QME)\cite{lindblad1976generators, breuer2002theory}, a differential equation of the density operator of a subsystem derived by reducing the freedom of reservoirs from the density operator of the total system and
using several approximations such as Markov reservoir. With QME,
many low-dimensional spin systems have been
studied\cite{vznidarivc2010matrix,prosen2011exact,buvca2016connected,prosen2015matrix,prosen2008third,medvedyeva2014power,vznidarivc2015relaxation,medvedyeva2016exact,shibata2019dissipative1,PhysRevB.99.224432}. Especially,
analysis has been carried out for steady state in
one-dimensional systems with the matrix ansatz\cite{vznidarivc2010matrix,prosen2011exact,buvca2016connected,prosen2015matrix}. In
these papers, the authors studied
one-dimensional open XX and XXZ spin chain in the steady state and obtained analytical
expressions for physical quantities such as correlation functions. For details, see a review\cite{prosen2015matrix}. In the case of transient dynamics,
analytical understanding is less satisfactory than the steady
case\cite{prosen2008third,medvedyeva2014power,vznidarivc2015relaxation,medvedyeva2016exact,shibata2019dissipative1,PhysRevB.99.224432}. In
\cite{vznidarivc2015relaxation}, the author studied the dependence of
the system size to relaxation time in XX and XXZ spin models with
dephasing. The spectrum of the Lindblad operator in several spin models with
dephasing has been studied\cite{medvedyeva2016exact,
shibata2019dissipative1, PhysRevB.99.224432}, but the analytical
understanding of transient dynamics with dissipation caused by reservoir has remained
to be investigated. 

The other method to explore the nonequilibrium state of open quantum system is the nonequilibrium Green function method(NEGF)\cite{keldysh1965diagram,schwinger1961brownian}. The
method is an extension of the standard Green function method so that one
can use the diagram techniques even for nonequilibrium systems. In
contrast to QME,  NEGF does not tell us information of the density
matrix itself about system which we focus on. However, understanding of
transient dynamics is relatively
progressing\cite{cini1980time,stefanucci2004time,
myohanen2009kadanoff,tuovinen2013time,ridley2015current,ridley2017partition,fukadai2018transient,karlsson2018generalized}. At
the beginning, NEGF did not include the effect of the initial
correlation between the system and reservoirs, which could not be ignored
when we focus on the transient dynamics. A further extension was
accomplished to include the initial
correlation in \cite{cini1980time,stefanucci2004time}. Using the formula,
dynamics of general systems has been studied
\cite{myohanen2009kadanoff,tuovinen2013time,ridley2015current,ridley2017partition}. In the
study\cite{myohanen2009kadanoff}, the authors paved the way for calculating the
Green functions including the effects of the Coulomb interaction using proper approximation of the self-energy to conserve physical
quantities such as energy. Recently, more efficient way of computing one-body
physical quantity has been used with the generalized Kadanoff-Baym ansatz\cite{latini2014charge,karlsson2018generalized}. Without the Coulomb interaction,
transport properties were calculated exactly\cite{tuovinen2013time} in the wide-band-limit approximation (WBLA). The method of calculation in
\cite{tuovinen2013time} has been employed to study the case where an
arbitrary time-dependent bias is applied to
reservoirs\cite{ridley2015current,ridley2017partition} and to
investigate double quantum dots\cite{fukadai2018transient}. However, the
application of general results in NEGF to concrete systems has not been enough.

In this paper, to obtain further understanding about dynamics of low-dimensional open quantum systems, we investigate the open tight-binding model coupled to two
reservoirs on its edges with the NEGF method including initial
effect. It is worth studying this model from two reasons. Theoretically,
dynamics of the isolated tight-binding model, or equivalently XX model which is obtained by Jordan-Wigner transformation, has been studied
extensively\cite{antal1999transport,ogata2002diffusion,hunyadi2004dynamic,platini2007relaxation}
and we can observe interesting phenomena such as scaling behavior of magnetization\cite{antal1999transport}. Therefore, we can expect further rich behaviors in the open case due to boundary effects. In addition, this model can be implemented experimentally by using cold atomic systems\cite{brantut2012conduction, sponselee2018dynamics}. Hence we can testify our results from experimental points of view. As a result, we obtain an analytical expression of the electron
density in the steady state and in the transient regime for $N
\rightarrow \infty$ and at absolute zero temperature, which we use to mimic the
situation of the experiment. From the expression of the electron density
in the steady state, we can prove a phase transition, whose transition point is the same as that obtained in
the isolated case\cite{sachdev1999quantum}, exists in
our open case. In addition, we show that the boundary
parameters do not change the qualitative behavior of the electron
density in the steady state. In contrast, we can see boundary effects in
the transient regime. We analytically understand that the
time-dependent electron density has a two-step decay in the region where the on-site
energy is not large enough and boundary couplings exist as
Fig. \ref{fig:rho_e} shows. This phenomenon
is purely caused by boundary parameters. We also show that the dependence of speed of
convergence to steady state on the boundary couplings changes at
a certain value (Fig. \ref{fig:phasetrans1}). It is reasonable to
consider that the speed of convergence increase as the couplings on the
boundaries become large because the couplings determine how easily
particles from reservoirs can flow into the sites. However, contrary to
the expectation, the
dependence actually changes at certain values of the coupling constants and the speed
decreases as the couplings become large. This is caused by special
eigenvalues, whose imaginary part does not vanish even in the limit $N
\rightarrow \infty$, of a non-hermitian matrix in the expression of the
electron density. Because the values of the boundary parameters determine the
existence of the special eigenvalues,
the qualitative change in the behavior of the time-dependent electron density is a result of the open boundaries.

This paper is organized as follows. In Sect. 2, we briefly review the
previous results obtained with the NEGF method including the initial
correlation effect. By considering the result in our case, we can obtain
a formal expression for the time-dependent electron density of site
$n$. In Sect. 3, we show several natures of the non-hermitian matrix, an important quantity of
one-body physical quantities. Especially, we investigate special
eigenvalues of the non-hermitian matrix. We need to pay attention to
the special eigenvalues when we analyze physical quantities because the eigenvalues change the expressions of the quantities. Based on the
understandings, we examine the behavior of the electron density in the
steady state and the transient regime in Sect. 4 and Sect. 5. In Sect. 4,
we study the electron density in steady state and obtain a simple
expression of the electron density. From the expression, we show the
existence of a phase transition and the boundary
parameters do not affect the qualitative behavior of the electron
density in the steady state. We investigate the time-dependent electron
density in Sect. 5. From the analytical expression of the time-dependent
electron density, we show the existence of a two-step decay and consider
the physical understanding of the behavior. We also find that the dependence of speed of convergence
to steady state on the boundary coupling constants changes at a certain
value. Sect. 6 is the conclusion of this study. 

\section{Our model and prevoirs results}
\begin{figure}[tb]
 \begin{center}
  \includegraphics[width=0.7\hsize]{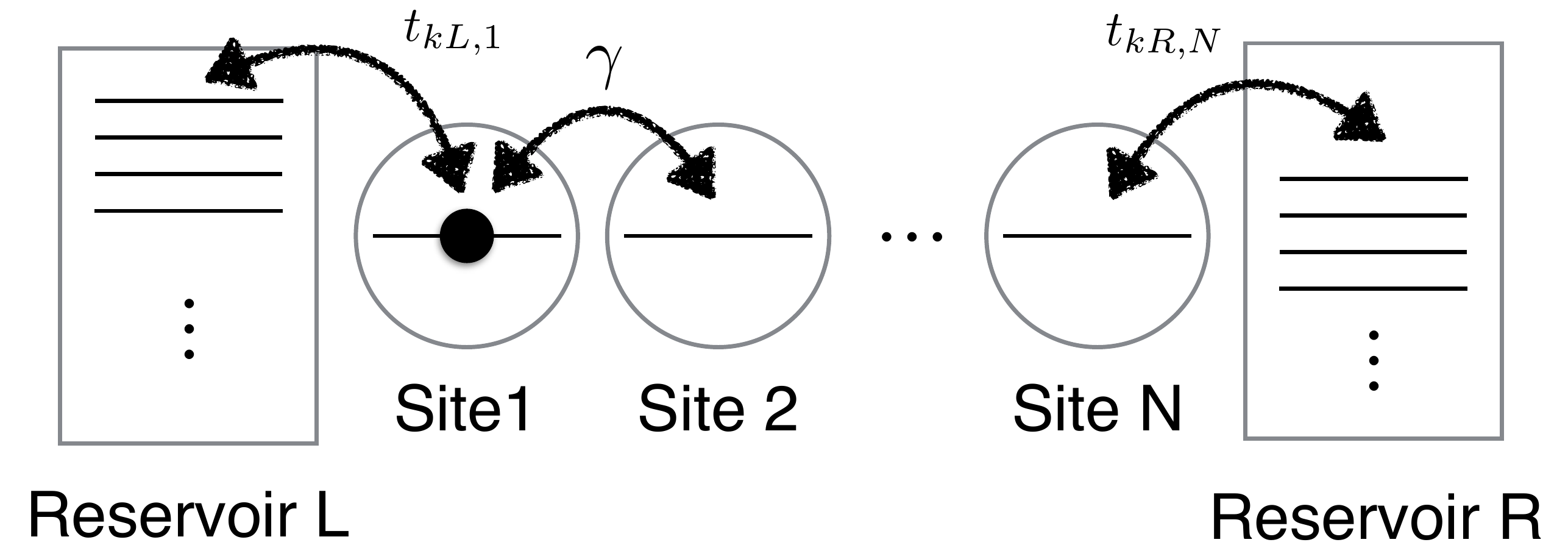}
\end{center}
\caption{Schematic diagram of our system.}
\label{fig:1}
\end{figure}
Our model is the straightly connected $N$ tight-binding model coupled to two
reservoirs(Fig. \ref{fig:1}). We consider the case where each site
has a single energy level. Therefore, a single electron can exist in each
site when we ignore the freedom of spin. The discussion can be extended
to the case including the freedom of spin straightforwardly. We assume that,
initially at time $t=0$, the sites and the reservoirs
are in the thermal-equilibrium state characterized by an inverse
temperature $\beta$ and a chemical potential $\mu$. Then for $t>0$, a constant bias voltage $V_{\alpha}$ is applied to
the reservoir $\alpha=\{\mathrm{L},\mathrm{R}\}$ and the system gets into a nonequilibrium
state. The Hamiltonian of total system with $N$ sites is represented as 
\begin{multline}
 H(t) = \\
\begin{cases}
  \sum_{\alpha=\{L,R\},k\alpha} (\epsilon_{k\alpha} - \mu ) d^{\dagger}_{k\alpha}
  d_{k\alpha} + \sum^N_{n=1} (\epsilon - \mu) d^{\dagger}_n d_n +
  V_{1L} + V_{NR} + \sum^{N-1}_{i=1} T_{n,n+1},  & t < t_0 \\
  \sum_{\alpha=\{L,R\},k\alpha} (\epsilon_{k\alpha} + V_{\alpha}) d^{\dagger}_{k\alpha}
  d_{k\alpha} + \sum^N_{n=1} \epsilon d^{\dagger}_n d_n +
  V_{1L} + V_{NR} + \sum^{N-1}_{n=1} T_{n,n+1}, & t \geq t_0
\end{cases} 
\label{eq:1} 
\end{multline}
where $\epsilon_{k \alpha}$ is the $k$th eigenvalue of the Hamiltonian
of reservoir $\alpha$ and $\epsilon$ is the on-site energy of the sites. $d_{k\alpha},d^{\dagger}_{k\alpha}$
are the creation and the annihilation operators of reservoir $\alpha$ and
$d_n,d^{\dagger}_n$ are those of site $n$.
The transported particles are Fermion and thus the operators satisfy
the anti-commutation relations: $\{d_i,d^{\dagger}_j\}=\delta_{ij}$,
$\{d_i,d_j\} = \{d^{\dagger}_i,d^{\dagger}_j\}=0$.
Each term of the Hamiltonian has the meaning as follows: the first term
is the (diagonalized) Hamiltonian of the reservoirs, the second is the
Hamiltonian of the sites, the third is the coupling between site 1 and the left reservoir,
the fourth is the coupling between site $N$ and the right reservoir. The
fifth is the coupling between the sites. Here, we assume the couplings
are written in the forms:
\begin{eqnarray*}
 V_{i \alpha} = \sum_{k\alpha} t_{k\alpha,i} d^{\dagger}_{k \alpha}
  d_i + t_{i, k\alpha} d^{\dagger}_i d_{k \alpha} \ (i=1,N), \\
  T_{n,n+1} = \gamma  d^{\dagger}_{n+1} d_n +  \gamma d^{\dagger}_{n} d_{n+1}\ (n
   =1,2,\cdots N-1).
\end{eqnarray*}
This means we consider the case
where Coulomb interaction is negligible. Experimentally, this approximation may be justified when the energy scale of electrons is larger than that of the Coulomb
interaction such as a high bias voltage.

In this situation, we want to know the quantitative behavior in the
steady state and the transient dynamics. To investigate this, we
calculate the electron density at site $i$ defined as
$\rho_{i}(t):=\braket{d^{\dagger}_{H,i}(t) d_{H,i}(t)}$. One of the methods to calculate physical quantities in nonequilibrium
state is NEGF. Until now, with the NEGF and WBLA, the expression of the Green function, which is the
basic quantity for one-body physical quantities, has been obtained in general situations without Coulomb interaction\cite{tuovinen2013time,ridley2015current}. Here WBLA means that the energy bands of
reservoirs are so wide and only the electrons on the Fermi
energy are transported. We use a formal expression of time-dependent electron
density $\rho_{n}(t)$ at site $n$ in \cite{tuovinen2013time} written as 
\begin{align}
 \rho_{n}(t) &= \big[ \sum_{\alpha=\{L,R \}} \int^{\infty}_{-\infty} \frac{d \omega}{2 \pi} f(\omega-\mu)
 \mathbf{\Lambda}_{\alpha}(\omega + V_{\alpha}) \notag \\
 &+ V_{\alpha} (e^{-i
 \mathbf{h}^{eff}(t-t_0)} \mathbf{G}^r(\omega) \mathbf{\Lambda}_{\alpha}(\omega +
 V_{\alpha}) e^{i(\omega+V_{\alpha})(t-t_0)} + \mathrm{H.c.}) \notag \\
 &+ V^2_{\alpha} e^{-i
 \mathbf{h}^{eff}(t-t_0)} \mathbf{G}^r(\omega) \mathbf{\Lambda}_{\alpha}(\omega +
 V_{\alpha}) \mathbf{G}^a(\omega) e^{i
 (\mathbf{h}^{eff})^*(t-t_0)} \big]_{nn},
\label{eq:2}
\end{align}
where $ \mathbf{\Lambda}_{\alpha} = \mathbf{G}^r(\omega) \Gamma_{\alpha}
\mathbf{G}^a(\omega)$ is a spectral function and $f(\omega)=\{1+ e^{\beta \omega}\}^{-1}$ is the Fermi
distribution function. In the following analysis, we set the chemical
potential $\mu$ to be 0 because $\mu$ is just a shift of energy. $\mathbf{G}^r(\omega)= \{\omega \mathbf{1}_{N} - \mathbf{h}^{eff}\}^{-1}$ and
$\mathbf{G}^a(\omega)= \{\omega \mathbf{1}_{N} - (\mathbf{h}^{eff})^*
\}^{-1}$ are the retarded Green function and the advanced Green function
respectively. Here $\mathbf{h}^{eff}= \mathbf{h} - \frac{i}{2} \mathbf{\Gamma}$
is a non-hermitian matrix including the effect from the reservoirs and representing the
geometrical configuration of the sites. Depending on the form of the
hamiltonian and couplings between sites and reservoirs, the expression
of $\mathbf{h}^{eff}$ changes. In our case of (\ref{eq:1}), each term of
the non-hermitian matrix is defined as follows. $\mathbf{\Gamma} = \sum_{\alpha} \mathbf{\Gamma}_{\alpha}$ and $\mathbf{\Gamma}_{\alpha}$ is
the matrix which expresses the coupling between the sites and the reservoir
$\alpha$. These matrices take the following forms in our case:
\begin{align}
 &\left[ \mathbf{\Gamma}_{L} \right]_{nm} = 
 \begin{cases}
 \Gamma_L, & n=m=1, \\
 0, & \mathrm{else},
 \end{cases}
 &\left[ \mathbf{\Gamma}_{R} \right]_{ij} = 
 \begin{cases}
 \Gamma_R, & i=j=N, \\
 0, & \mathrm{else}.
 \end{cases}
\label{eq:3}
\end{align}
The expression of the matrix $\mathbf{\Gamma}_{\alpha}$ changes depending on the couplings between the sites and reservoirs. Because the left reservoir only connects with the site 1 in our case,
$\mathbf{\Gamma}_{L}$ has only the non-zero value $\Gamma_L$ at $(1,1)$
component. Similarly, $\mathbf{\Gamma}_{R}$ only has the non-zero value
$\Gamma_R$ at $(N,N)$ component. We note that the parameters $t_{k
\alpha, n}$ in (\ref{eq:1}) representing the effect of the reservoir
$\alpha$ are summarized into the single matrix $\mathbf{\Gamma}_{\alpha}$ after the WBLA. $\mathbf{h}$ is the
hermitian matrix of the system and its form is written as
\begin{equation}
 \left[ \mathbf{h} \right]_{nm} = 
 \begin{cases}
 \epsilon, & n=m, \\
 \gamma, & |n-m|=1, \\
 0, & \mathrm{else}.
 \end{cases}
\label{eq:4}
\end{equation} 
The expression of the matrix $\mathbf{h}$ changes depending on the
coupling between the sites. Because each site is connected only to its
nearest neighbors in our case, the coupling constant $\gamma$ only exists at $|n-m|=1$. 
From the expression (\ref{eq:2}), we need to diagonalize the
non-hermitian matrix $\mathbf{h}^{eff}$ when we proceed with the analysis of the electron
density (\ref{eq:2}).

\section{Analysis of the non-hermitian matrix $\mathbf{h}^{eff}$}
\begin{figure}[tb]
 \begin{center}
  \includegraphics[width=0.5\hsize]{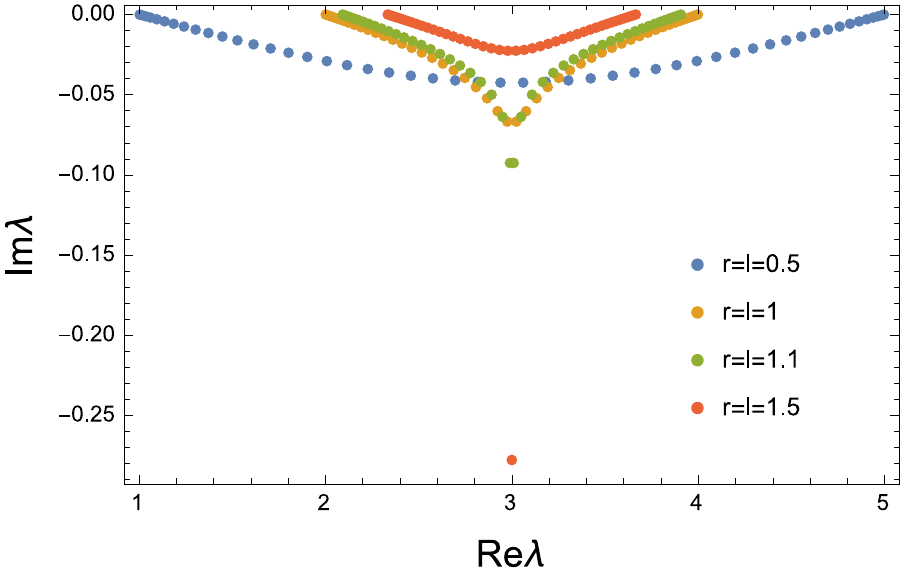}
\end{center}
\caption{(Color Online) The distribution  of the
 eigenvalues $\lambda$ for $N=50, \epsilon=3, \Gamma_L=\Gamma_R=1$. The horizontal
 axis is the real part of the eigenvalue. We change the
 parameter $\gamma$ to satisfy $r=l=0.5,\ 1,\ 1.1,\ 1.5$. We can see
 that the special eigenvalues appear in $r=l>1$.}
\label{fig:2}
\end{figure}

\begin{figure}[tb]
\begin{minipage}[t]{0.45\hsize}
 \begin{center}
  \includegraphics[width=0.95\hsize]{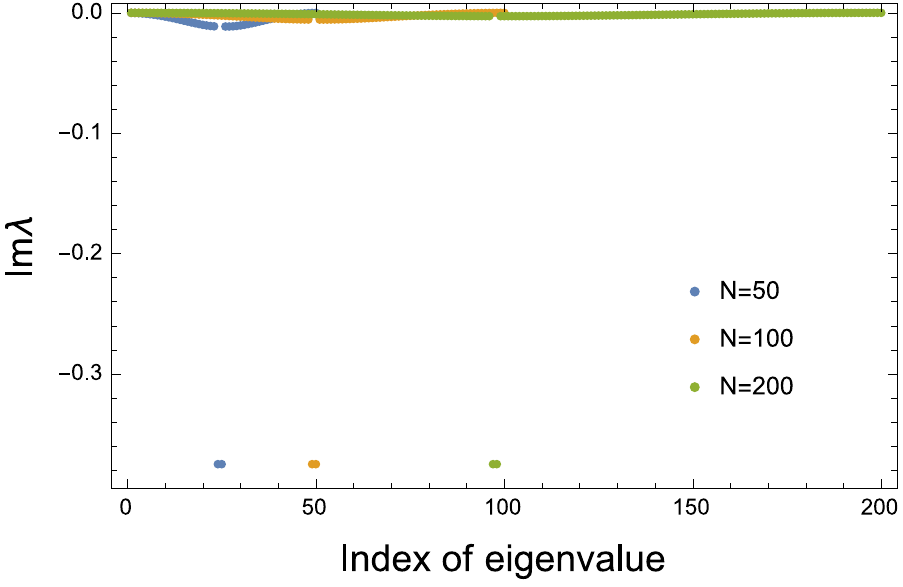}
\end{center}
\caption{(Color Online) The distribution of the imaginary parts of the
 eigenvalues for $\epsilon=3, \Gamma_L=\Gamma_R=1, \gamma=0.25$ as a function of the label of the eigenvalue. We consider the
 cases: $N=50, 100, 200$. For all numbers of sites, the imaginary part of the special eigenvalue takes the same value.}
\label{fig:3}
\end{minipage}
\hfil
\begin{minipage}[t]{0.45\hsize}
\begin{center}
 \includegraphics[width=0.95\hsize]{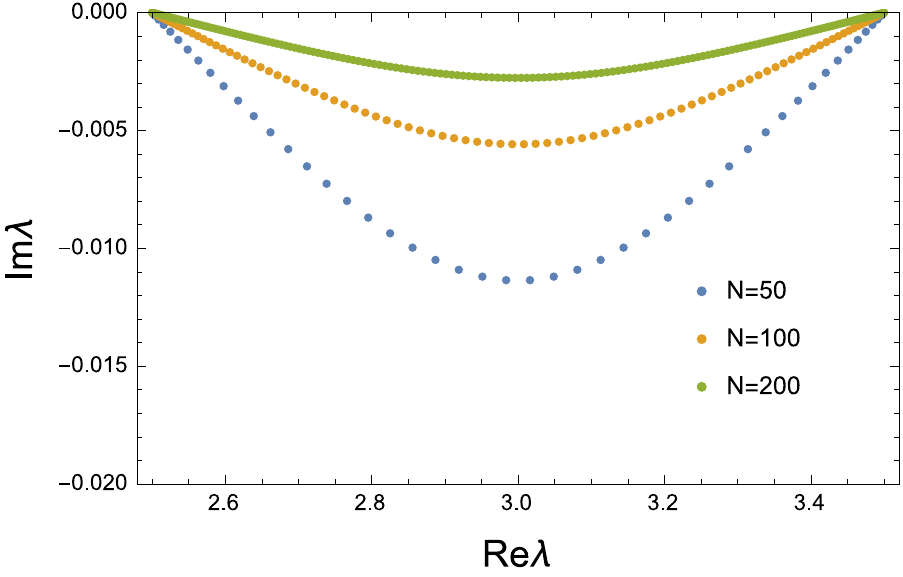}
\end{center}
\caption{(Color Online) The distribution of the imaginary parts of the normal
 eigenvalues for $\epsilon=3, \Gamma_L=\Gamma_R=1, \gamma=0.25$. We consider the
 cases: $N=50, 100, 200$. We can see that imaginary parts of the normal eigenvalues decrease as $O(\frac{1}{N})$.}
\label{fig:4}
\end{minipage}
\end{figure}

In this section, we investigate the non-hermitian matrix
$\mathbf{h}^{eff}$ which appears
in the expression for the time-dependent electron density (\ref{eq:2})
and show several properties of the matrix, which we use later for the analysis of the electron density in Sect. 4 and Sect. 5.
The non-hermitian matrix which we focus on is 
\begin{equation}
 \mathbf{h}^{eff} = 
\begin{pmatrix}
     \epsilon - \frac{i}{2} \Gamma_{L} & \gamma & 0 & \cdots & & & &\\
     \gamma & \epsilon & \gamma & 0 & \cdots & & & \\
     0 & \gamma & \epsilon & \gamma & \cdots & & & \\
     \vdots & & & &0  & \gamma  & \epsilon & \gamma \\
     0 & \cdots & & & & 0 & \gamma & \epsilon - \frac{i}{2} \Gamma_{R} \\
 \end{pmatrix}.
\label{eq:5}
\end{equation}
First we note that this matrix can have special
eigenvalues whose imaginary parts do not vanish even in the
thermodynamics limit $N \rightarrow \infty$. Imaginary parts of
other eigenvalues vanish as $O(\frac{1}{N})$. In the case of $l=\Gamma_L/2\gamma<1$ and $r=\Gamma_R/2\gamma<1$, there is no special
eigenvalue. When $l>1$ or $r>1$ is satisfied, there is one special
eigenvalue. In the region where $l,r>1$, there are two special
eigenvalues. We need to pay attention to the fact when we analyze
physical quantities because the special eigenvalues change the
expression of the physical quantities. In the transient case, not only the
expression but also the qualitative behavior of the electron density
changes as we see in Sect. 5. We can check the existence of the special
eigenvalues and the necessary condition of their appearance as follows. We express an eigenvalue of the non-hermitian matrix as
$\lambda$. We can obtain the characteristic equation of the non-hermitian matrix as 
\begin{gather}
(\beta^{N+1} - \alpha^{N+1}) + i (l+r) (\beta^{N} - \alpha^{N}) - lr(\beta^{N-1} - \alpha^{N-1}) = 0, \label{eq:6} \\
 \alpha + \beta = \tilde{\lambda},\ \alpha \beta = 1, \label{eq:7} 
\end{gather}
where we define the normalized eigenvalue
$\tilde{\lambda}=(\lambda-\epsilon)/\gamma$. For a derivation of (\ref{eq:6}), see Appendix A. By considering the limit $N \rightarrow \infty $ of the
characteristic equation, we can show the existence of the special
eigenvalues as follows. We assume that $N$ is
sufficiently large. When we look for solutions which satisfy
$|\beta| > 1$, which is equivalent to $|\alpha|<1$ from the condition $\alpha
\beta=1$ in Eq. (\ref{eq:7}), we can ignore the term $\alpha^N$ in (\ref{eq:6}). Then, by dividing by $\beta^N$, Eq. (\ref{eq:6}) becomes 
\begin{equation*}
\beta^2 + i(l+r)  \beta - lr = 0.
\end{equation*}
The solutions of this equation are $\beta=-ir, -il$, which corresponds
to the special eigenvalues. Note that each solution exists only when the
parameter satisfy $l>1$ and $r>1$ respectively to satisfy the assumption
$|\beta|=r,l>1$. Therefore, the solutions of
Eq. (\ref{eq:6}) in $N \rightarrow \infty$ which
satisfy $|\beta|>1$ only exist when $r>1$ or $l>1$ and these are
$\beta=-il, -ir$. From Eq. (\ref{eq:7}), the corresponding eigenvalues are
\begin{align}
 \lambda &= \epsilon -i \gamma (l - \frac{1}{l}),\ \epsilon -i \gamma (r -
  \frac{1}{r})
\label{eq:8} \\
 &=: \Lambda_l,\ \Lambda_r, \notag
\end{align}
where we define the special eigenvalue related to the parameters of the left reservoir and the
right reservoir as $\Lambda_l$ and $\Lambda_r$ respectively. In a
similar manner, we consider solutions of the case $|\beta| < 1$. The only
solutions of Eq. (\ref{eq:6}) in this case are $\beta=i/l$ and $\alpha=i/r$ with the condition $l
>1$ and $r>1$ respectively. The corresponding eigenvalues are $\Lambda_l$ and $\Lambda_r$, the same as in the case
$|\beta| > 1$. Therefore, the remaining solutions should satisfy
$|\beta| = 1$, or $\beta = e^{i \theta}$, $\theta \in \mathcal{R}$. This means that the imaginary part of the other
eigenvalues are 0 in the $N \rightarrow \infty$ limit since other eigenvalues are expressed as $\lambda
= \epsilon + 2 \gamma \cos \theta$ from Eq. (\ref{eq:7}). We numerically check these facts obtained above. Fig. \ref{fig:2} is the distribution of the eigenvalues for
several $r=l=\Gamma/2 \gamma$ obtained by diagonalizing the non-hermitian matrix
$\mathbf{h}^{eff}$ numerically.  From Fig. \ref{fig:2}, we can see that whether the
special eigenvalues appear or not is determined by the values of the
parameters $r,l$. When the parameters satisfy $r=l>1$, the special eigenvalues appear.
Fig. \ref{fig:3} shows the distribution of the eigenvalues for
several numbers of the sites $N$. From this graph, the imaginary parts of
the special eigenvalues do not disappear and do not change their values
even in the large $N$ limit. In contrast, the imaginary
parts of other eigenvalues decrease as the number of the sites $N$
increases shown in Fig. \ref{fig:4}. Fig. \ref{fig:4} is the
graph of the distribution of the normal eigenvalues for several $N$. The
imaginary parts of the normal eigenvalues gradually decrease with
$O(1/N)$ as the
number of the sites increases\cite{vznidarivc2015relaxation}. 

Until now, we study the eigenvalues in limit $N \rightarrow
\infty$. Next we consider the case of finite $N$, where we show a symmetry of
the distribution of the eigenvalues for analysis of the electron density
in the steady state and see an expression of the
eigenvectors. First we obtain another expression of the characteristic
equation (\ref{eq:6}). We express the LHS of Eq. (\ref{eq:6}) as
$\Delta_N(\tilde{\lambda})$. By using
the identity $\beta^N - \alpha^N = (\beta + \alpha)(\beta^{N-1} -
\alpha^{N-1}) - \alpha \beta (\beta^{N-2} - \alpha^{N-2})$, we can show that
$\Delta_{N}(\tilde{\lambda})$ satisfies the following recurrence relation
\begin{equation}
 \Delta_{N}(\tilde{\lambda}) = \tilde{\lambda}
  \Delta_{N-1}(\tilde{\lambda}) - \Delta_{N-2}(\tilde{\lambda}),
\label{eq:9}
\end{equation}
for $N \geq 2$ with the initial conditions $\Delta_{0}(\tilde{\lambda})
= lr+1$ and $\Delta_{1}(\tilde{\lambda}) =
\tilde{\lambda}+i(l+r)$. For derivation, we use $\alpha \beta =1$ from (\ref{eq:7}). Polynomials which satisfy the recurrence
relation (\ref{eq:9}) are called the modified Chebyshev polynomials\cite{witula2006modified}. It
is known\cite{witula2006modified} that any modified Chebyshev polynomial $g_{N}(z)$, $z \in \mathbf{C}$ is expressed as
\begin{equation}
 g_{N}(z) = g_1(z) V_{N-1}(z) - g_0(z) V_{N-2}(z),
\end{equation}
where $V_N(z)=U_N(z/2)$ is the modified Chebyshev polynomial of the
second kind and $U_N(\cos \Theta) = \sin((N+1) \Theta) / \sin(\Theta)$, $\Theta \in \mathcal{C}$, is the Chebyshev polynomial of the first kind. Using this property in our case $g_N = \Delta_N$, we obtain the expression $\Delta_{N}(\tilde{\lambda}) = (\tilde{\lambda} +
i(l+r))V_{N-1}(\tilde{\lambda}) - (lr+1) V_{N-2}(\tilde{\lambda})$. By changing the
variable $\tilde{\lambda} = 2 \cos \Theta$, we finally arrive at another
expression for the characteristic equation\cite{yueh2005eigenvalues}
\begin{gather}
(2 \mathrm{cos} \Theta + i(l+r)) \mathrm{sin} N \Theta - (lr +1)
 \mathrm{sin} (N-1) \Theta = 0, \label{eq:10} \\
 \tilde{\lambda} =2 \mathrm{cos} \Theta. \notag
\end{gather}
Eq. (\ref{eq:10}) has all the information about the eigenvalues. From this expression, we can show a symmetry about the distribution of the eigenvalues $\tilde{\lambda}_k$. When Eq. (\ref{eq:10}) has a solution $\tilde{\lambda}_k$, there is the
corresponding eigenvalue $- (\tilde{\lambda}_k)^*$, which we call the
conjugate eigenvalue of $\tilde{\lambda}_k$. We can prove this as
follows. By dividing Eq. (\ref{eq:10}) by $\sin
\Theta$ and applying the recurrence relation $2 \tilde{\lambda}
U_N(\tilde{\lambda}) -U_{N-1}(\tilde{\lambda}) =
U_{N+1}(\tilde{\lambda})$, we have the expression,
\begin{equation}
 U_{N+1}(\tilde{\lambda}_k) +i(l+r)  U_N(\tilde{\lambda}_k) -lr
  U_{N-1}(\tilde{\lambda}_k) = 0.
\label{eq:10.1}
\end{equation}
This expression can also be derived from Eq. (\ref{eq:7}) by setting $\alpha = e^{i \Theta_k}$. By taking the complex conjugate of Eq. (\ref{eq:10.1}), we have
\begin{align*}
 (U_{N+1}(\tilde{\lambda}_k) +i(l+r)  U_N(\tilde{\lambda}_k) - lr 
  U_{N-1}(\tilde{\lambda}_k) )^*
&= U_{N+1}(\tilde{\lambda}^*_k) - i(l+r) U_N(\tilde{\lambda}^*_k) - lr
 U_{N-1}(\tilde{\lambda}^*_k) \notag \\
&=  (-1)^{N+1} U_{N+1}(- \tilde{\lambda}^*_k) - (-1)^{N} i(l+r) U_N(-\tilde{\lambda}^*_k) \\ 
& \qquad - lr (-1)^{N-1} U_{N-1}(-\tilde{\lambda}^*_k) \notag \\
&= U_{N+1}(- \tilde{\lambda}^*_k) + i(l+r)
 U_N(-\tilde{\lambda}^*_k) - lr U_{N-1}(-\tilde{\lambda}^*_k) \\
&= 0.
\end{align*}
From the first line to the second line, we use the relation about the
Chebyshev polynomial of the second kind $U_N(-z) = (-1)^N U_N(z)$. This
result shows that $- (\tilde{\lambda}_k)^*$ is also the eigenvalue when
$\tilde{\lambda}_k$ is the eigenvalue. We can see this from
Fig. \ref{fig:2}. Except for the case where $\lambda_k$ is a pure imaginary
number, there is always the conjugate eigenvalue $-
(\tilde{\lambda}_k)^*$ of the eigenvalue $\lambda_k$.

Finally we discuss the eigenvector $\mathbf{r}_k$ corresponding to the eigenvalue
$\lambda_k$. Because of the fact $(\mathbf{h}^{eff})^t=\mathbf{h}^{eff}$, the
following relation about the eigenvectors holds 
\begin{equation}
 \sum_{k} \frac{\mathbf{r}_k \mathbf{r}^t_k}{\mathbf{r}^t_k \cdot
  \mathbf{r}_k}  = \sum_k \mathbf{R}_k =1,
\label{eq:10.4}
\end{equation}
where we define $\mathbf{R}_k = \frac{\mathbf{r}_k \mathbf{r}^t_k}{\mathbf{r}^t_k \cdot
  \mathbf{r}_k}$. The concrete expression for the eigenvector is written as\cite{yueh2005eigenvalues}
\begin{equation}
 \mathbf{r}^{(n)}_k = \sin n \Theta_k + i l \sin (n-1) \Theta_k.
\label{eq:10.2}
\end{equation}
Especially, the eigenvector $\mathbf{r}_{\Lambda_l}$ for the special eigenvalue, 
$\Lambda_l$, is written as
\begin{equation}
  \mathbf{r}^{(n)}_{\Lambda_l} = - \frac{(-il)^{-n+1}}{2}(l+\frac{1}{l}).
\label{eq:10.2.1}
\end{equation}
From these expression above, $[\mathbf{R}_k]_{nm}$ and $[\mathbf{R}_{\Lambda_l}]_{nm}$, which is $\mathbf{R}_k$ of the special eigenvalue $\Lambda_l$, are expressed as 
\begin{gather}
 [\mathbf{R}_k]_{nm} \eqsim \frac{2}{N} \frac{(\sin n \Theta_k + i l \sin
  (n-1) \Theta_k)(\sin m \Theta_k + i l \sin
  (m-1) \Theta_k)}{1-l^2 + 2 i l \cos \Theta_k},
\label{eq:10.3.1} \\
 [\mathbf{R}_{\Lambda_l}]_{nm} \eqsim - (1 + l^2) (- i l)^{-(n+m)}.
\label{eq:10.3.2}
\end{gather}
We prove Eq. (\ref{eq:10.2.1}) to Eq. (\ref{eq:10.3.2}) in Appendix A. Here $\eqsim$ means that we ignore the order of
$O(1/N)$. 
The symmetry of the eigenvalues is written in terms of the eigenvectors as follows. Since we can express the conjugate eigenvalue as $- (\tilde{\lambda}_k)^* = - 2 \cos (\Theta_k)^* = 2 \cos ( \Theta^*_k +
\pi)$, we can obtain the corresponding eigenvector of the conjugate
eigenvalue by adding $\pi$ to the phase $\Theta^*_k$ in the eigenvector
(\ref{eq:10.2}). The result is 
\begin{align}
\sin j ( \Theta^*_k + \pi) + i l \sin (j-1) ( \Theta^*_k  + \pi) 
&= (-1)^j (\sin j \Theta^*_k - i l  \sin (j-1) \Theta^*_k ) \notag \\
&= (-1)^j (\mathbf{r}^{j}_k)^*.
\label{eq:10.3} 
\end{align}
We use this relation to study the electron density in the steady state.

\section{Steady state}
In this and the next section, we investigate the behavior of the
electron density in our open system. First we consider the case of steady
state and investigate boundary effects, especially the effects of the
special eigenvalues, to the electron density. To carry out
it, we derive a simple expression for the electron density in the steady
state for general number of site $N$ and absolute zero temperature case
$\beta = \infty$. We need this manipulation because we do not have an
concrete expression of the eigenvalues themselves for general $N$ as we
see in the previous section and therefore cannot know the qualitative
behavior of the electron
density in the steady state from the original representation (\ref{eq:14}). 
From the simple expression of the electron density in the steady state,
first we show that a phase
transition, whose behavior is the same as
that obtained in the isolated case\cite{sachdev1999quantum}, also exists in our open case. In
addition, we can analytically derive $\rho^{ss}_n =
\frac{1}{2}$ for $\epsilon = V = 0$ with the expression and the symmetric
distribution of the eigenvalues. These behavior has been seen in the
analysis of QME\cite{prosen2008quantum, vznidarivc2010matrix},
but the formulation is different from our model. About effects of the
special eigenvalues, we confirm that the special eigenvalues do not
change the qualitative behavior of the electron density in the steady
state for thermodynamic limit $N \rightarrow \infty$.

From the expression for the time-dependent electron density (\ref{eq:2}), we can obtain the expression for the electron density in the steady state $\rho^{ss}$ as
\begin{equation}
 \mathbf{\rho}^{ss}  =  \int^{\infty}_{-\infty} \frac{d \omega}{2 \pi}
 \sum_{\alpha} f(\omega) \mathbf{\Lambda}_{\alpha}(\omega + V_{\alpha}), 
\label{eq:12}
\end{equation}
whose $n$th diagonal element expresses the electron density at site
$n$. This is because the second and the third term in (\ref{eq:2}) have
the term $e^{-i \mathbf{h}^{eff} (t-t_0)}$ which vanishes in the limit
$t \rightarrow \infty$ as the imaginary parts of the eigenvalues of $\mathbf{h}^{eff}$ are negative. To know the behavior of the electron density
(\ref{eq:12}) in detail, we rewrite the expression (\ref{eq:12}) using the
eigenvalues $\lambda_k$ and the eigenvectors $\mathbf{r}_k$ of $\mathbf{h}^{eff}$. 
By using the relation (\ref{eq:10.4}), we have the following expression
\begin{align}
 \mathbf{\rho}^{ss} &= \sum_{\alpha} \sum_{k,l} \left[  \mathbf{R}_k  \mathbf{\Gamma}_{\alpha}  \mathbf{R}^*_l  \right] \int^{\infty}_{-\infty} \frac{d \omega}{2 \pi}
 f(\omega-\mu) \frac{1}{(\omega + V_{\alpha} -
 \lambda_k) (\omega + V_{\alpha} - \lambda^*_l)} \notag \\
&=  \sum_{\alpha} \sum_{k,l} \left[  \mathbf{R}_k  \mathbf{\Gamma}_{\alpha} \mathbf{R}^*_l  \right] \frac{1}{\lambda_k - \lambda^*_l}
 \left( F_1(\lambda_k- V_{\alpha}) - F_1(\lambda^*_l- V_{\alpha} ) \right), \label{eq:14}
\end{align}
where we define $ F_1(\lambda) =
\int^{\infty}_{-\infty} \frac{d \omega}{2 \pi} f(\omega-\mu)  \frac{1}{\omega- \lambda}$. We note that all the effects of temperature $\beta$
are included in $F_1(\lambda)$. In the case of $\beta=\infty$, we can write the analytical expression for $F_1(\lambda)$ in (\ref{eq:14}) as 
\begin{align}
 F_1(\lambda_k- V_{\alpha}) - F_1(\lambda^*_l - V_{\alpha}) &= \frac{1}{2 \pi} \left[
 \log(\omega - (\lambda_k - V_{\alpha})) -  \log(\omega - (\lambda^*_l - V_{\alpha}))
 \right]^{0}_{-\infty} \notag \\
&= \frac{1}{2 \pi} (\log( V_{\alpha} -\lambda_k) - \log( V_{\alpha} - \lambda^*_l) - 2 \pi i), \label{eq:14.1.1}
\end{align}
where we use the fact $\mathrm{Im} (\lambda_k)<0$. By substituting the
expression (\ref{eq:14.1.1}) into (\ref{eq:14}), we have
\begin{align}
\mathbf{\rho}^{ss}
&= \frac{1}{2 \pi} \sum_{\alpha} \sum_{k,l}  \left[  \mathbf{R}_k  \mathbf{\Gamma}_{\alpha} \left(
 \mathbf{R}_l \right)^* \right] \frac{1}{\lambda_k - \lambda^*_l}
 (\log( V_{\alpha} -\lambda_k) - \log( V_{\alpha} -\lambda^*_l) - 2 \pi i) \notag \\
&= 1 + \frac{1}{2 \pi} \sum_{\alpha} \sum_{k,l}  \left[  \mathbf{R}_k  \mathbf{\Gamma}_{\alpha} \left(
 \mathbf{R}_l \right)^* \right] \frac{1}{\lambda_k - \lambda^*_l}
 (\log( V_{\alpha}-\lambda_k) - \log( V_{\alpha} -\lambda^*_l)), 
\label{eq:14.1}
\end{align}
where we use the fact $ -i \sum_{k,l} \mathbf{R}_k  \mathbf{\Gamma} 
 \mathbf{R}^*_l / (\lambda_k - \lambda^*_l)= \mathbf{1}$ proved in Appendix B. We change the expression (\ref{eq:14.1}) into the more convenient form to discuss the behavior of the electron density in the steady state. By rearranging the expression
(\ref{eq:14.1}) as in Appendix B, we have the following representation for any $V_L$ and $V_R$
\begin{align}
\mathbf{\rho}^{ss} &= 1 + \frac{i}{2\pi} \sum_{\alpha} (
 \log( V_{\alpha} - \mathbf{h}^{eff} ) - \log( V_{\alpha} - (\mathbf{h}^{eff})^* )) \notag \\
& \quad - \frac{1}{2 \pi} \sum_{k,l} \mathbf{R}_k \frac{1}{\lambda_k
 - \lambda^*_l} \bigl( ( \log(V_L - \lambda_k ) - \log ( V_L - \lambda^*_l) )
 \mathbf{\Gamma}_{R} \notag \\
& \qquad +  ( \log(V_R - \lambda_k ) - \log ( V_R - \lambda^*_l))
 \mathbf{\Gamma}_{L}  \bigr)  \left( \mathbf{R}_l \right)^*.
\label{eq:rhost}
\end{align}
In addition, we assume symmetric bias voltages $V_L=V_R=V$ for simplicity. From the assumption, the electron
density (\ref{eq:rhost}) is written as
\begin{align}
 \mathbf{\rho}^{ss}
&= 1 + \frac{i}{\pi}  (
 \log( V - \mathbf{h}^{eff} ) - \log( V - (\mathbf{h}^{eff})^* )) \notag \\
& \quad - \frac{1}{2 \pi} \sum_{k,l} \mathbf{R}_k \mathbf{\Gamma} \left(
 \mathbf{R}_l \right)^* \frac{1}{\lambda_k
 - \lambda^*_l} \bigl( \log (V - \lambda_k) - \log (V - \lambda^*_l)
 \bigr) \notag \\
&= 1 + \frac{i}{\pi}  (
 \log( V - \mathbf{h}^{eff} ) - \log( V- (\mathbf{h}^{eff})^* )) -
 (\rho^{ss} -1),
\end{align}
where we use the definition of $\rho^{ss}$, ({\ref{eq:14.1}}) to obtain the final line. Therefore, the electron density takes the simple form
\begin{align}
  \mathbf{\rho}^{ss}
&= 1 + \frac{i}{2 \pi}  (
 \log( V - \mathbf{h}^{eff} ) - \log( V - (\mathbf{h}^{eff})^* ))
 \notag \\
&=  1 + \frac{i}{2 \pi}( \sum_{k} \mathbf{R}_k \log(V -
 \lambda_k) - \mathbf{R}^*_k \log(V - \lambda^*_k) ) \notag \\
&= 1 - \frac{1}{\pi} \mathrm{Im} \left[ \sum_{k} \mathbf{R}_k \log(V -
 \lambda_k) \right] \notag \\
&= 1 - \frac{1}{\pi} \mathrm{Im}  \left[ \log( V - \mathbf{h}^{eff} ) \right].
\label{eq:10.5}
\end{align}
With the expression (\ref{eq:10.5}), we discuss the behavior of the electron density in the steady state for
the case $N \rightarrow \infty$. Without loss of
generality, we set $V=0$ in the discussion since $V$ is just a shift of the energy. 

\begin{figure}[tb]
 \begin{center}
  \includegraphics[width=0.5\hsize]{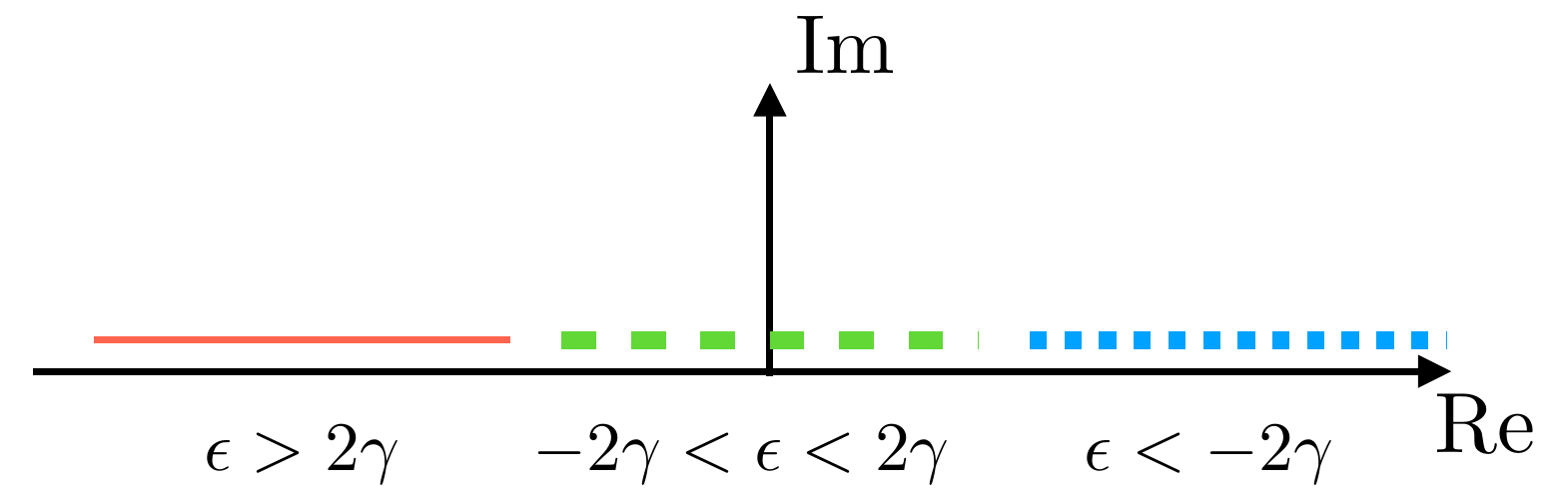}
\end{center}
\caption{(Color Online) Distribution of $- \lambda_k$ for large $N$. For
 the case $\epsilon > 2 \gamma$(solid line), all the eigenvalues are on the
 negative real line, whose argument are $\pi$. In contrast, the argument of
 all eigenvalues are $0$ for the case $ \epsilon < - 2 \gamma$ since
 they are on the positive real line(dots line). When $ - 2 \gamma < \epsilon < 2
 \gamma$ is satisfied, some eigenvalues are on the negative real
 line and the others are on the positive real line(dashed line).}
\label{fig:eigendis}
\end{figure}

\begin{figure}[tb]
 \begin{center}
  \includegraphics[width=0.5\hsize]{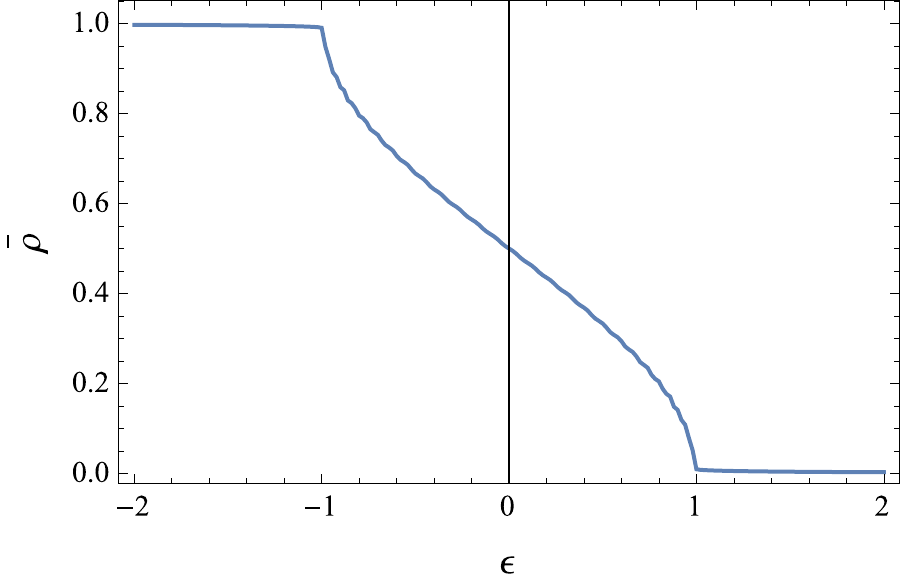}
\end{center}
\caption{(Color Online) The average electron denisty $\bar{\rho}$ and
 the on-site energy $\epsilon$. The parameters are set as $N=50,\ V_L=V_R=0,\
 \Gamma_L=\Gamma_R=\gamma=0.5$. The average electron density
 discontinuously changes at $\left| \frac{\epsilon}{2 \gamma} \right|=1$.}
\label{fig:avegsteady_e}
\end{figure}

First we investigate the average electron density per site $\bar{\rho}=\frac{1}{N} \mathrm{Tr} \mathbf{\rho}^{ss}$. By taking the
trace of Eq.(\ref{eq:10.5}) and dividing by $N$, we have the following expression
\begin{align}
\bar{\rho} 
&= 1 - \frac{1}{N} \frac{1}{\pi} \mathrm{Im}  \left[ \mathrm{Tr} \log( -
 \mathbf{h}^{eff} ) \right] \notag \\
&= 1 - \frac{1}{N} \frac{1}{\pi} \mathrm{Im}  \left[  \log( \mathrm{det}( -
 \mathbf{h}^{eff} ) ) \right] \notag \\
&= 1 - \frac{1}{N} \frac{1}{\pi} \mathrm{Im}  \left[ \log( \prod_{k} (-
 \lambda_k)) \right] \notag \\ 
&= 1 - \frac{1}{N} \frac{1}{\pi} \sum_{k} \mathrm{arg}(- \lambda_k).
\label{eq:10.6}
\end{align}
From the first to the second line, we use $\mathrm{Tr} \log(-\mathbf{h}^{eff}) = \log
\mathrm{det}(- \mathbf{h}^{eff})$. From this expression, we can
show that a phase transition exits in our open system as follows. First we consider the case $r,l<1$ for simple discussion, where there is no special
eigenvalue. We discuss the case with special eigenvalues later. In the
following discussion, we use the fact that the eigenvalues are expressed
as $\lambda_k \eqsim \epsilon + 2 \gamma \cos k + i0$, $k\in(0,\pi)$ for large $N$ as we show in Section 3. $\eqsim$ means that we ignore the terms of order $O(1/N)$. With this fact and the expression (\ref{eq:10.6}), we can see that there are three situations depending on the parameter
$\epsilon$ and $\gamma$. See the Fig. \ref{fig:eigendis}.
In the case $\epsilon > 2 \gamma$, $- \lambda_k$ are always on the
negative real line. Therefore, $\mathrm{arg}(- \lambda_k) = \pi$ holds and the average electron
density per site is $\bar{\rho} = 0$ from (\ref{eq:10.6}). When $ - 2 \gamma < \epsilon < 2
\gamma$ is satisfied, we set $N^*$ as the number of eigenvalues which
satisfy $\epsilon + 2 \gamma \cos k < 0$. Then the average electron density
is expressed as $\bar{\rho} = 1 - N^* /N$. In the limit $ N \rightarrow \infty$, we can easily prove the behavior is arccosine from Eq. (\ref{eq:10.6}) and the result is compatible with the numerical result (Fig. \ref{fig:avegsteady_e}). For the case of $ \epsilon < -
2 \gamma$, all the eigenvalues are on the positive real line and the
argument $\mathrm{arg}(- \lambda_k)$ is 0. Therefore, the average
electron density is $\bar{\rho} = 1$. In this way, we can see that the
phase transition exits in the open case
at $\left| \frac{\epsilon}{2 \gamma} \right|=1$ and the behavior is the
same result obtained in the isolated case\cite{sachdev1999quantum}. Actually, we can numerically show the phase
transition as in Fig. \ref{fig:avegsteady_e} obtained by computing the
definition of the electron density in the steady state (\ref{eq:14}).
We can physically understand this behavior as follows. When the on-site
energy $\epsilon$ is too high, which corresponds to the case of $\epsilon >
2 \gamma$, all electrons outflow from the sites to the reservoirs and
the average density finally becomes 0 in the steady state. In contrast,
electrons are accumulated until the sites are filled with electrons for
the case of $\epsilon > - 2 \gamma$. Therefore, the average electron
density is 1 in the steady state. When the same bias voltages $V$ are
applied to the reservoirs, the effect is just shifting the graph in Fig. \ref{fig:avegsteady_e} along the
horizontal line. Next we consider the effect of the special eigenvalues
to the behavior of the average electron density. Actually, the special
eigenvalue does not affect the average electron density in the limit $N \rightarrow \infty$. We can understand
this as follows. Let us consider the case $l<1, r>1$ where there is one special
eigenvalues $\Lambda_r$. In this case, the average electron density
(\ref{eq:10.6}) for large $N$ is written as
\begin{equation}
 \bar{\rho} = 1 - \frac{1}{N} \frac{1}{\pi} \left( \sum_{k: \mathrm{normal}} \mathrm{arg}(-
  \lambda_k) + \mathrm{arg}(-
  \Lambda_r) \right).
\label{eq:10.7}
\end{equation}
From the Eq. (\ref{eq:10.7}), we can see that the term from the special
eigenvalue vanishes in the $N \rightarrow \infty$ since the order of $\mathrm{arg}(-
  \Lambda_r)$ is $O(N^0)$. Therefore, the average electron density with the
special eigenvalue in $N \rightarrow \infty$ behaves in a similar way as the case without special eigenvalue. 

We also discuss the electron density at each site. First we analytically prove that the electron density at any site takes
the same value $\rho^{ss}_n = \frac{1}{2}$ for the case $\epsilon=V=0$. A similar situation is
considered with the different approach of QME\cite{vznidarivc2010matrix}
and the result is the same. After that, we numerically investigate the
dependence of the electron density at each site on the boundary
parameters. As a result, we see that the qualitative behavior of the electron
density in the steady state does not change by the special eigenvalues. For simplicity, we consider the case where the number of the sites
is even $N=2M$. In the case $\epsilon=0$, the eigenvalue is expressed
for $\lambda_k = \gamma \tilde{\lambda}_k$ from the definition (\ref{eq:7}). From the Eq. (\ref{eq:10.5}), we have 
\begin{align}
 \rho^{ss}_n &= 1 - \frac{1}{\pi} \mathrm{Im}  \sum^N_{k=1} \left[ \mathbf{R}_k \right]_{nn}
 \log( - \lambda_k)  \notag \\
 &= 1 - \frac{1}{\pi} \mathrm{Im}  \sum^M_{k=1} \left[  [ \mathbf{R}_k ]_{nn}
 \log( - \gamma \tilde{\lambda}_k) + [ \mathbf{R}^*_k ]_{jj}
 \log(  \gamma \tilde{\lambda}^*_k) \right],  \notag 
\end{align}
where we use the symmetric distribution of the normalized eigenvalue,
which means that there are always the corresponding eigenvalue $-
\tilde{\lambda}^*_k$ and the matrix $(-1)^{n+m}[\mathbf{R}^*_k]_{nm}$ to
the eigenvalue $\tilde{\lambda}_k$ and the matrix $[\mathbf{R}_k]_{nm}$. We can prove this
fact from (\ref{eq:10.3}). With the relations $ |\gamma
\tilde{\lambda}^*_k | = | - \gamma
\tilde{\lambda}_k |$ and $\mathrm{arg}(- \gamma \tilde{\lambda}_k) = \pi
- \mathrm{arg}( \gamma \tilde{\lambda}^*_k)$, we can express $\rho^{ss}_n$ as 
\begin{align}
  \rho^{ss}_n &= 1 - \frac{1}{\pi} \mathrm{Im}  \sum^M_{k=1} \bigg[  [ \mathbf{R}_k ]_{nn}
 ( \log | - \gamma \tilde{\lambda}_k | + i \mathrm{arg}(- \gamma
 \tilde{\lambda}_k) ) + [
 \mathbf{R}^*_k ]_{nn} ( \log | - \gamma \tilde{\lambda}_k | + i ( \pi -
 \mathrm{arg}( - \gamma \tilde{\lambda}_k) )  \bigg]  \notag \\
&= 1 - \mathrm{Im}  \sum^M_{k=1} i [
 \mathbf{R}^*_k ]_{nn}  \notag \\
&= 1 - \mathrm{Re}  \sum^M_{k=1}  [
 \mathbf{R}^*_k ]_{nn}.  \notag
\end{align}
Again we use the fact of the symmetric distribution of the normalized
eigenvalue and the eigenvector. Finally, we can write the electron density
$\rho^{ss}_n$ for $\epsilon = 0 $ as
\begin{align}
\rho^{ss}_n &= 1 - \frac{1}{2} \sum^M_{k=1}  [
 \mathbf{R}_k ]_{nn} + [ \mathbf{R}^*_k ]_{nn}  \notag \\
&= 1 - \frac{1}{2} \sum^N_{k=1}  [
 \mathbf{R}_k ]_{nn} \notag \\
&= \frac{1}{2}.
\label{eq:10.8}
\end{align}
To obtain the final result, we use Eq. (\ref{eq:10.4}). 
\begin{figure}[tb]
\begin{minipage}[t]{0.45\hsize}
 \begin{center}
  \includegraphics[width=0.95\hsize]{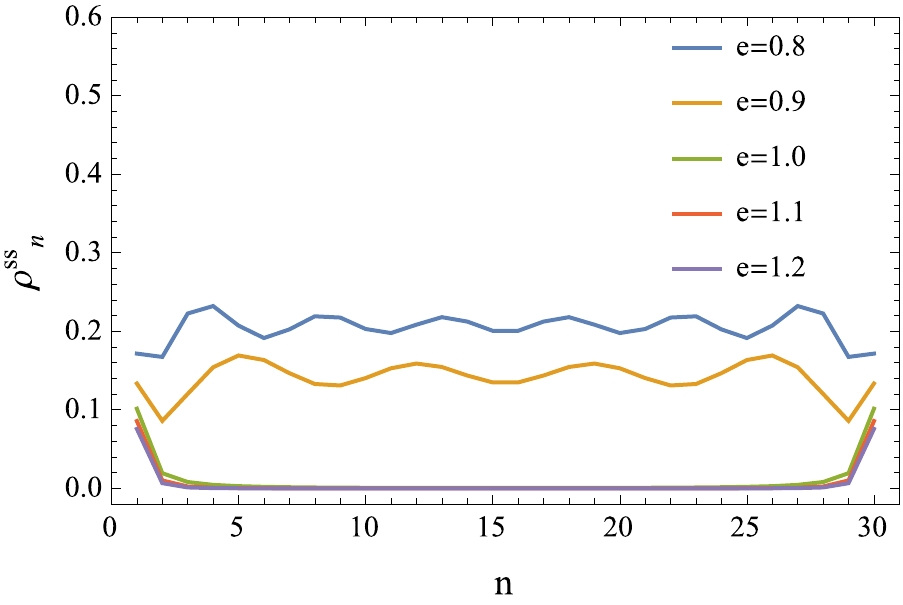}
\end{center}
\caption{(Color Online) The electron denisty $\rho_n$ for several $\epsilon$. The parameters are set as $N=30,\ V_L=V_R=0,\
 \Gamma_L=\Gamma_R=\gamma=0.5$. The electron density at each site
 discontinuously changes at $\epsilon =  2 \gamma = 1$}
\label{fig:eachsteady_e1}
\end{minipage}
\hfil
\begin{minipage}[t]{0.45\hsize}
\begin{center}
 \includegraphics[width=0.95\hsize]{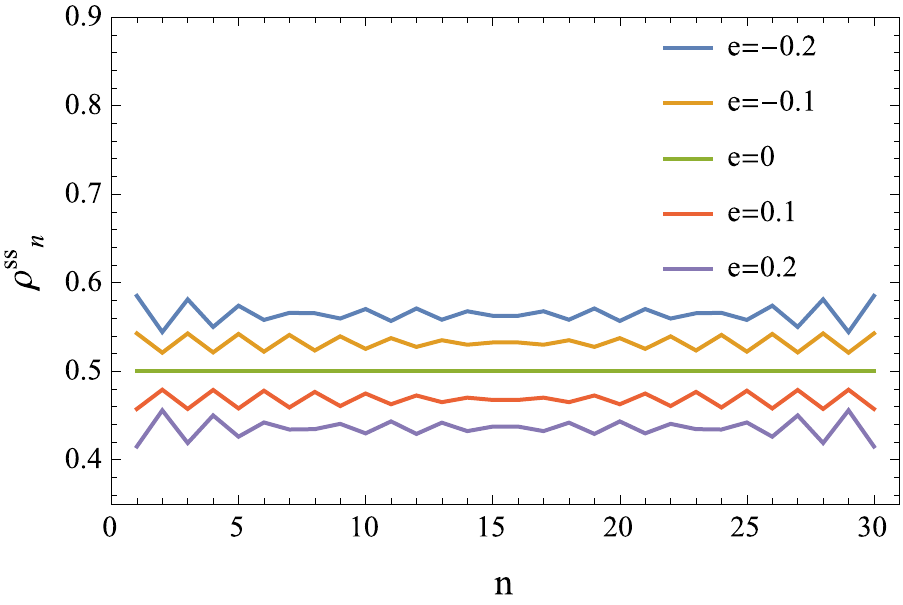}
\end{center}
\caption{(Color Online) The electron denisty $\rho_n$ for $-2 \gamma <
 \epsilon < 2 \gamma$. The parameters are set as $N=30,\ V_L=V_R=0,\ \Gamma_L=\Gamma_R=\gamma=0.5$. In this region, the electron density at each site fluctuates.}
\label{fig:eachsteady_e2}
\end{minipage}
\end{figure}
\begin{figure}[tb]
\begin{minipage}[ht]{0.45\hsize}
 \begin{center}
  \includegraphics[width=0.95\hsize]{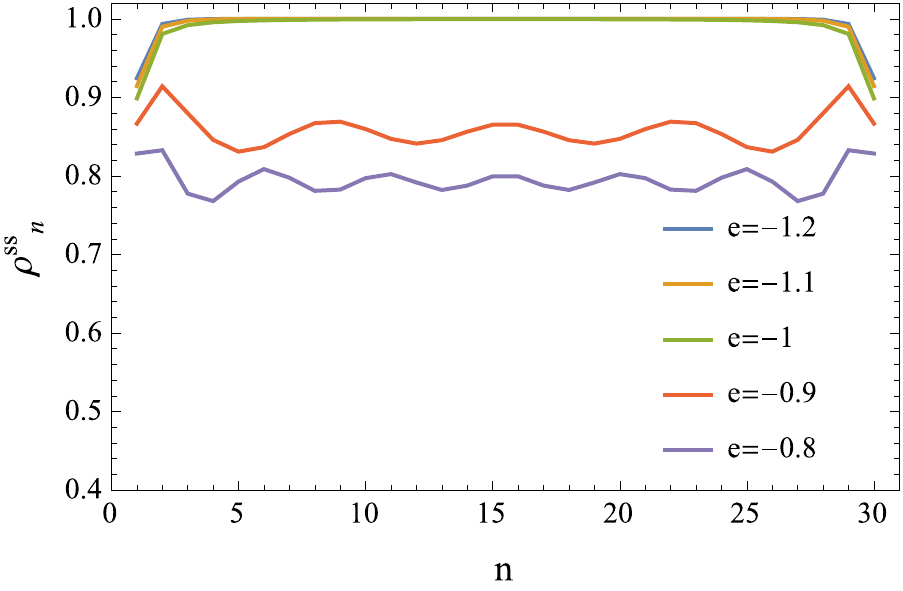}
\end{center}
\caption{(Color Online) The electron denisty $\rho_n$ for several $\epsilon$. The parameters are set as $N=30,\ V_L=V_R=0,\
 \Gamma_L=\Gamma_R=\gamma=0.5$. The electron density at each site
 discontinuously changes at $\epsilon = - 2 \gamma = -1$}
\label{fig:eachsteady_e3}
\end{minipage}
\hfil
\begin{minipage}[ht]{0.45\hsize}
\begin{center}
 \includegraphics[width=0.95\hsize]{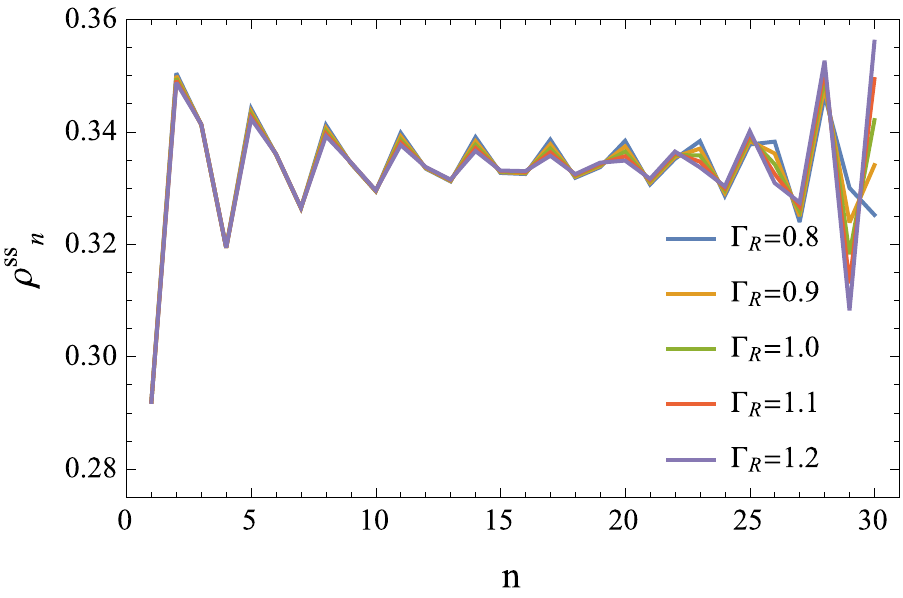}
\end{center}
\caption{(Color Online) The electron denisty $\rho_n$ for several $\Gamma_R$. The parameters are set as $N=30,\ V_L=V_R=0,\
 ,\epsilon=\Gamma_L=\gamma=0.5$. We cannot see any discontinuous behavior at $r=\frac{\Gamma_R}{2\gamma}=1$.}
\label{fig:eachsteady_g}
\end{minipage}
\end{figure}
We numerically calculate the electron density at site $n$ in the steady state from the definition (\ref{eq:14}) for several energy $\epsilon$. The results are Fig. \ref{fig:eachsteady_e1}, Fig. \ref{fig:eachsteady_e2} and Fig. \ref{fig:eachsteady_e3}. These figures show that the
phase transition at $|\frac{\epsilon}{2 \gamma}|=1$ also exists not only in the average electron density
but also in the electron density at each site except the sites on the edges. We call the sites not the left most site and the right most site as the bulk sites for explanation. From
Fig. \ref{fig:eachsteady_e1}, we can see that the electron density is
almost 0 on the bulk sites for the case $\epsilon > 2 \gamma$. When the
parameters satisfy $\epsilon < -2 \gamma$, the electron density on the
bulk sites are almost 1 as Fig. \ref{fig:eachsteady_e3} shows. The value
of the electron density suddenly changes at $\epsilon = 2 \gamma$ and
$\epsilon = - 2 \gamma$. The behavior is the same as that of the average
electron density. For the case where $ -2 \gamma < \epsilon < 2\gamma$,
there is a difference from the average electron density(Fig. \ref{fig:eachsteady_e2}). The electron density fluctuates by sites around the
average electron density except $\epsilon = 0$. At $\epsilon = 0$, the
electron density does not depend on the site index and it takes the same
value $\rho^{ss}_n= \frac{1}{2}$, which we have shown analytically (\ref{eq:10.8}). The
fluctuation is suppressed as the on-site energy $\epsilon$ is close to
$0$. We also study the dependence of $\rho^{ss}_n$ on the coupling
parameter at the right boundary $\Gamma_R$ to see the
effect of the special eigenvalue(Fig. \ref{fig:eachsteady_g}). In the settings, the special eigenvalue appears at $\Gamma_L =1$. However, we
cannot see any discontinuous behavior at $r=1$ from
Fig. \ref{fig:eachsteady_g}. In conclusion, the phase transition
of the electron density in the steady state, whose behavior is the
same as that of the isolated tight-binding model, also exists in our
open case. The boundary terms do not break the phase transition and do
not change the qualitative behavior of the electron density by the special eigenvalue.

\section{Time-dependent case}

\begin{figure}[tb]
 \begin{center}
  \includegraphics[width=0.55\hsize]{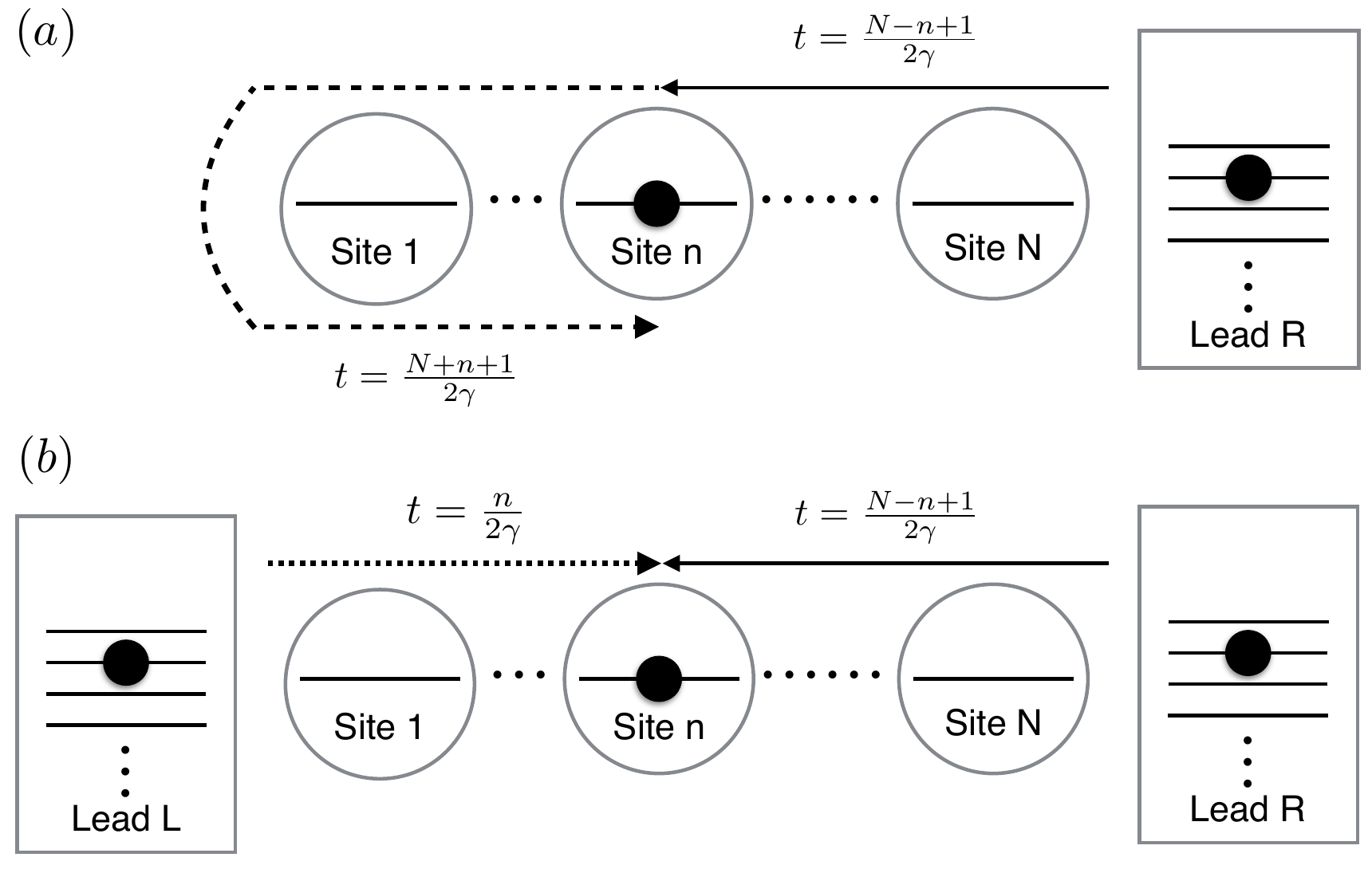}
\end{center}
\caption{Schematic diagram of transport for (a) $l=0,\
 r\neq0$ and (b) $l \neq 0,\ r \neq 0$. In the case $l=0$, Electrons from the
 left reservoir reach site $n$ around $t=\frac{N+1-n}{2\gamma}$(Real line)
 for the first time. After reflection at the right edge, the electrons
 come back to site $n$ around $t=\frac{N+1+n}{2 \gamma}$(Dashed line). There are
 electrons from the right reservoir in the case $l \neq 0$, which comes
 to site $n$ around $t=n$(Dotted Line) or $t=\frac{N-n+1}{2\gamma}$(Real line).}
\label{fig:dg_transport}
\end{figure}

\begin{figure}[tb]
\begin{minipage}[t]{0.45\hsize}
 \begin{center}
  \includegraphics[width=0.9\hsize]{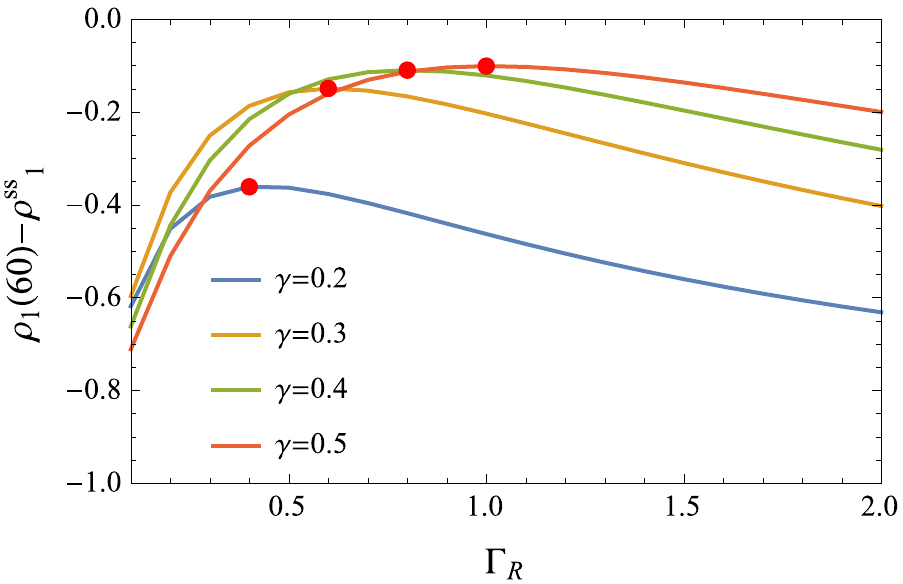}
\end{center}
\caption{(Color Online) The behavior of the time-dependent electron
 density at a time $\rho_1(60) - \rho^{ss}_1$ for several $\gamma$ and
 $\Gamma_R$. We set the parametrs as $N=20,\ V_L=V_R=6,\
 \epsilon=3, \ \Gamma_L=0$. Depending on the parameters $\gamma$, the peak(red point)
 changes and is compatible with out expectation $\Gamma_R
 = 2 \gamma$ for all the cases.}
\label{fig:phasetrans1}
\end{minipage}
\hfil
\begin{minipage}[t]{0.45\hsize}
\begin{center}
 \includegraphics[width=0.9\hsize]{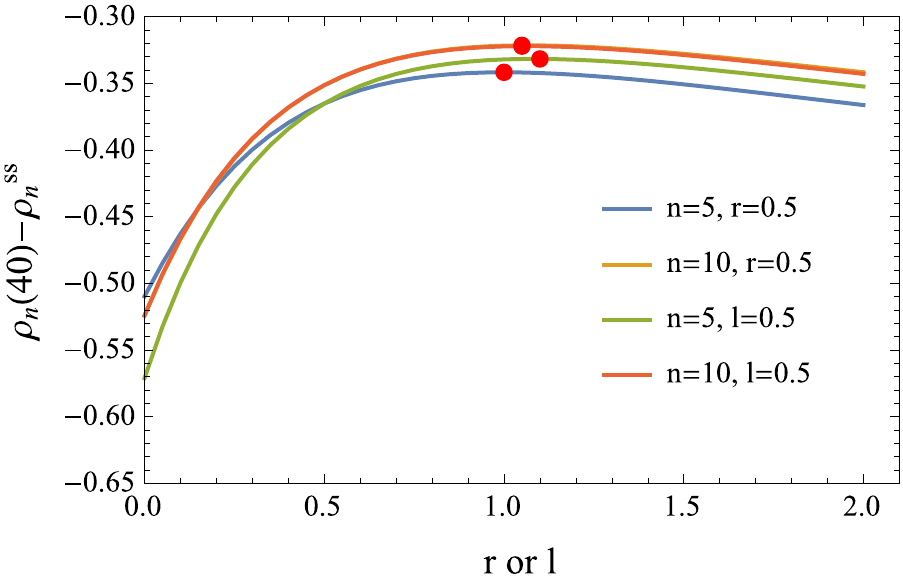}
\end{center}
\caption{(Color Online) The behavior of the time-dependent electron
 density $\rho_n(t) - \rho^{ss}_n$ for several site $n$ by coupling parameter $l$ or $r$. We
 fix $l$ when we change the parameters $r$ and vice versa. We set the
 parametrs as $N=20,\
 V_L=V_R=6,\ \gamma=0.5, \epsilon=3$. We consider $t=40$ as a different time from that in Fig \ref{fig:phasetrans1}.}
\label{fig:phasetrans2}
\end{minipage}
\end{figure}

Next we consider the time-dependent case, which is of our main
interest and investigate the effect of the boundary parameters. With the calculation technique used also in the analysis of
steady state, we derive an analytical expression of the time-dependent electron density for $N \rightarrow \infty$ and zero
temperature case $\beta = \infty$. From the expression, first we see that a two-step
relaxation appears in the time-dependent electron density caused by boundary couplings for the case of $\epsilon
< V + 2 \gamma$, which cannot be seen in the previous study about the dynamics of
isolated case\cite{antal1999transport}. Because of the boundary couplings, particles flow and the time lag between flow of
particles from the left and the right reservoir causes the two-step
decay, which we understand later using
Fig. \ref{fig:dg_transport}. For the case of $\epsilon \geq V + 2
\gamma$, the time-dependent electron density decays to 0 quickly. This
is because the on-site energy of the dots are too high and electrons
cannot remain in the dots. Next we study the effects of the special
eigenvalues to the time-dependent electron density. By computing the analytical expression of
the electron density for the case with special eigenvalues, we find that the
dependence of speed of convergence to steady state on the boundary
couplings changes by the special eigenvalue. We can expect that the speed of
convergence increase as the couplings on the boundaries become large because the boundary couplings determine how easily
particles from reservoirs can flow into the sites. However, the
dependence changes by the special eigenvalues at $l=1$ and $r=1$ and the speed decrease as the
couplings become large. We check this fact from numerical calculation of the definition of the time-dependent electron density
(Fig. \ref{fig:phasetrans1}).

We start our analysis from the formal expression of the electron density (\ref{eq:2}). The time-dependent part of the electron density is expressed as 
\begin{align}
 \rho(t) - \rho^{ss} &=  \sum_{\alpha=\{L,R \}} \int^{\infty}_{-\infty}  \frac{d \omega}{2
\pi} f(\omega-\mu) V_{\alpha} (e^{-i
 \mathbf{h}^{eff}(t-t_0)} \mathbf{G}^r(\omega) \mathbf{\Lambda}_{\alpha}(\omega +
 V_{\alpha}) e^{i(\omega+V_{\alpha})(t-t_0)} + \mathrm{H.c.}) \notag \\
 & \quad + V^2_{\alpha} e^{-i
 \mathbf{h}^{eff}(t-t_0)} \mathbf{G}^r(\omega) \mathbf{\Lambda}_{\alpha}(\omega +
 V_{\alpha}) \mathbf{G}^a(\omega) e^{i
 (\mathbf{h}^{eff})^*(t-t_0)} \notag \\
 &= \sum_{\alpha=\{L,R \}} 2 \mathrm{Re} [\mathbf{\rho}^{(1)}_{\alpha}(t)] + \mathbf{\rho}^{(2)}_{\alpha}(t).
\label{eq:16}
\end{align}
First we investigate the behavior of
$\mathbf{\rho}^{(1)}_{\alpha}(t)$. This is because 
$\mathbf{\rho}^{(2)}_{\alpha}(t)$ behaves in almost the same manner as
$\mathbf{\rho}^{(1)}_{\alpha}(t)$ as we see later. For simplicity, we
assume a symmetric and large bias voltages: $V_L=V_R=V$ and $V \gg |\lambda_k -
\lambda^*_l| \eqsim 4 \gamma$. We study other cases using numerical
calculation. By using the assumption of the symmetric bias voltages and carrying out the same calculation as in the derivation of (\ref{eq:rhost}), we have the following expression
\begin{align}
\mathbf{\rho}^{(1)}(t) 
&=\sum_{\alpha=\{L,R \}} \mathbf{\rho}^{(1)}_{\alpha}(t) \notag \\
&= i \sum_{k} \mathbf{R}_k e^{- i (\lambda_k - V) t} (F_2(t, \lambda_k)
 - F_2(t, \lambda_k - V) ) \notag \\
& \quad -  i \sum_{k,l} \mathbf{R}_k \mathbf{R}^*_l \frac{V}{\lambda_k -
 \lambda^*_l + V} e^{- i (\lambda_k - V)t} (F_2(t, \lambda_k) - F_2(t,
 \lambda^*_l - V) ), 
\label{eq:17}
\end{align}
where we define $F_2(t, z) = \int^{\infty}_{-\infty} \frac{d \omega}{2
\pi} f(\omega- \mu) \frac{e^{ i \omega t}}{\omega - z}$. $F_2(t,z)$ includes all the effects of the temperature. We can express $F_2(t,z)$ for $\beta=\infty$ as 
\begin{equation}
 F_2(t,z) = \frac{e^{i t z}}{2 \pi} \times
\begin{cases}
 \left( - E_1(itz) + 2 \pi i \right), & \mathrm{Re}z <0\ \mathrm{and}\ \mathrm{Im}z >0, \\
 - E_1(itz), & \mathrm{else},
\end{cases}
\label{eq:18}
\end{equation}
where $E_n(x)=z^{n-1} \int^{\infty}_z \frac{e^{-t}}{t^{n}}$ is the $n$th
order of the exponential integral with its principal value. The
derivation is in Appendix C. We note that $F_2(t,z)$ is a multivalued function. The two cases in the expression (\ref{eq:18}) arises due to the branch cut of the exponential function $E_1(z)$ on the negative
real line. At this point, we use the assumption of the large bias voltages $V \gg 4 \gamma$. From this assumption, the relation $\frac{V}{\lambda_k -
 \lambda^*_l + V} \eqsim 1$ holds, where we ignore the order of $O(\frac{1}{V})$. By using the relation, we can rewrite (\ref{eq:17}) in a simple form as
\begin{align}
 \mathbf{\rho}^{(1)}(t) 
&= i \sum_{k} \mathbf{R}_k e^{- i (\lambda_k - V) t} (F_2(t, \lambda_k)
 - F_2(t, \lambda_k - V) ) \notag \\
& \quad -  i \sum_{k,l} \mathbf{R}_k \mathbf{R}^*_l e^{- i (\lambda_k - V)t} (F_2(t, \lambda_k) - F_2(t,
 \lambda^*_l - V) ) \notag \\
&= - i \sum_{k} \mathbf{R}_k e^{- i (\lambda_k - V) t} 
 F_2(t, \lambda_k - V) + i  \sum_{k,l} \mathbf{R}_k \mathbf{R}^*_l
 e^{- i (\lambda_k - V)t} F_2(t, \lambda^*_l - V) \notag \\
&= \mathbf{\rho}^{(1,1)}(t) +  \mathbf{\rho}^{(1,2)}(t),
\label{eq:19}
\end{align}
where we define 
\begin{gather}
 \mathbf{\rho}^{(1,1)}(t) = - i \sum_{k} \mathbf{R}_k e^{- i (\lambda_k - V) t} 
 F_2(t, \lambda_k - V), \label{eq:19.1}  \\
\mathbf{\rho}^{(1,2)}(t) =  i  \sum_{k,l} \mathbf{R}_k \mathbf{R}^*_l
 e^{- i (\lambda_k - V)t} F_2(t, \lambda^*_l - V). \label{eq:19.2}
\end{gather}
We investigate the behavior of (\ref{eq:19.1}) and (\ref{eq:19.2}) in
detail. As we will see, both quantities take different forms depending
on the existence of the special eigenvalues, $\Lambda_l$ and $\Lambda_r$. From this fact, the dependence of convergence speed on the boundary parameters $l,r$ changes at $l=1$ and $r=1$.  We also note that the term
$\mathbf{\rho}^{(1,2)}(t)$ shows two-step decay for the case of
$\epsilon < V + 2 \gamma$ and $l>0$ or $r>0$. In other words,
the quantity have quasi-steady state if the on-site energy is not large
enough to satisfy $\epsilon < V + 2 \gamma$ and there is a coupling
between the edge dots and the reservoirs. This is because the
argument of $F(t, \lambda^*_l -
V)$ in $\mathbf{\rho}^{(1,2)}(t)$ crosses the branch cut on the negative
real line for the case $\epsilon < V + 2 \gamma$ and a phase term
arises. $\mathbf{\rho}^{(1,1)}(t)$ does not have the behavior of
two-step decay because the argument of $F(t, \lambda_k -V)$ never cross
the branch cut. We check these facts below.

\begin{figure}[tb]
\begin{minipage}[t]{0.45\hsize}
 \begin{center}
  \includegraphics[width=0.95\hsize]{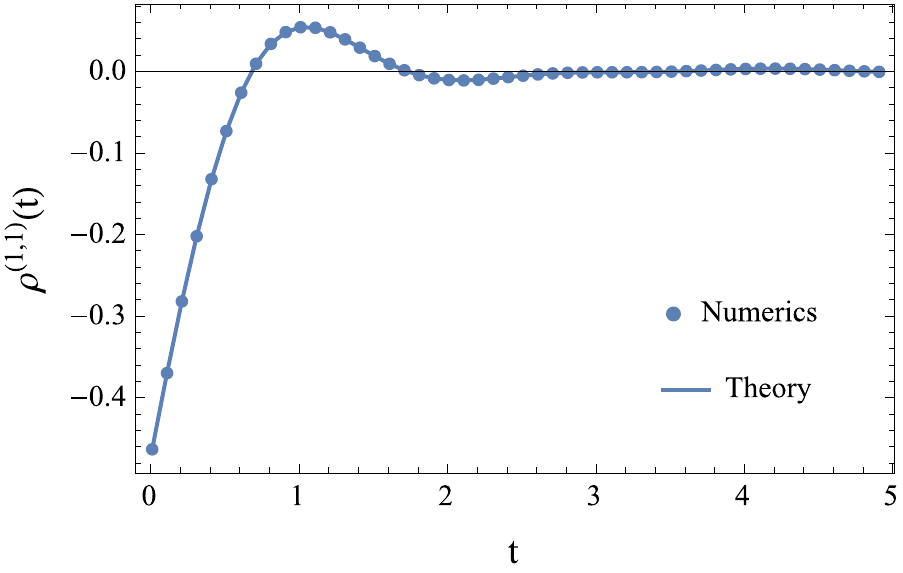}
\end{center}
\caption{(Color Online) The behavior of the quantity $\mathbf{\rho}^{(1,1)}_1(t)$ obtained
 numericaly(Blue dots) and analytically(Blue line). We set the
 parameters as $N=20,\ V_L=V_R=6,\ \epsilon=3,\ \Gamma_R=\gamma=0.5$. We can see the
 good agreement of the analytical result with the numerical result.}
\label{fig:rho11}
\end{minipage}
\hfil
\begin{minipage}[t]{0.45\hsize}
\begin{center}
 \includegraphics[width=0.95\hsize]{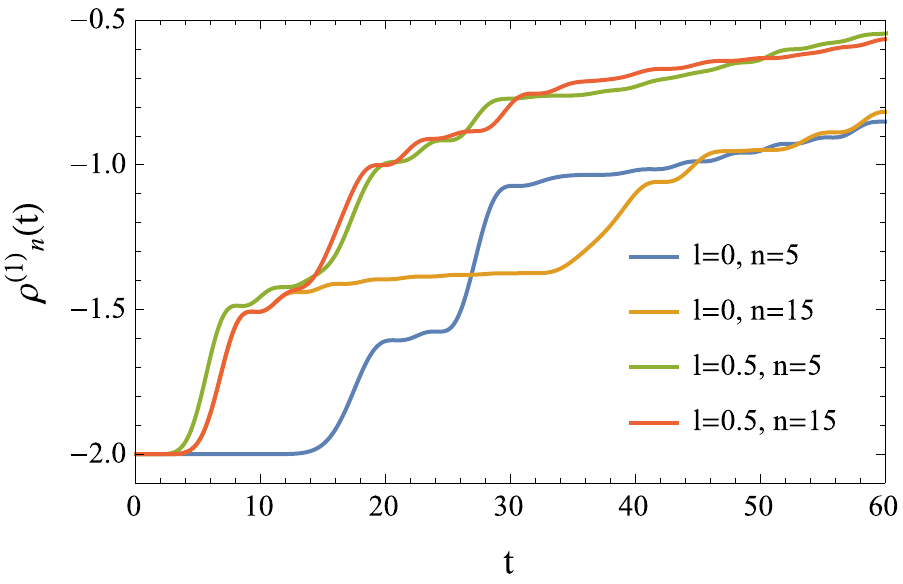}
\end{center}
\caption{(Color Online) The behavior of the quantity $\rho^{(1)}_n(t)$ by
 the site index $n$ and the coupling parameter $l$. We set the parameters as $N=20,\
 V_L=V_R=6,\ \epsilon=3,\ \Gamma_R=\gamma=0.5$. We can see two-step decay.}
\label{fig:rho1}
\end{minipage}
\end{figure}

\begin{figure}[tb]
\begin{minipage}[t]{0.45\hsize}
 \begin{center}
  \includegraphics[width=0.95\hsize]{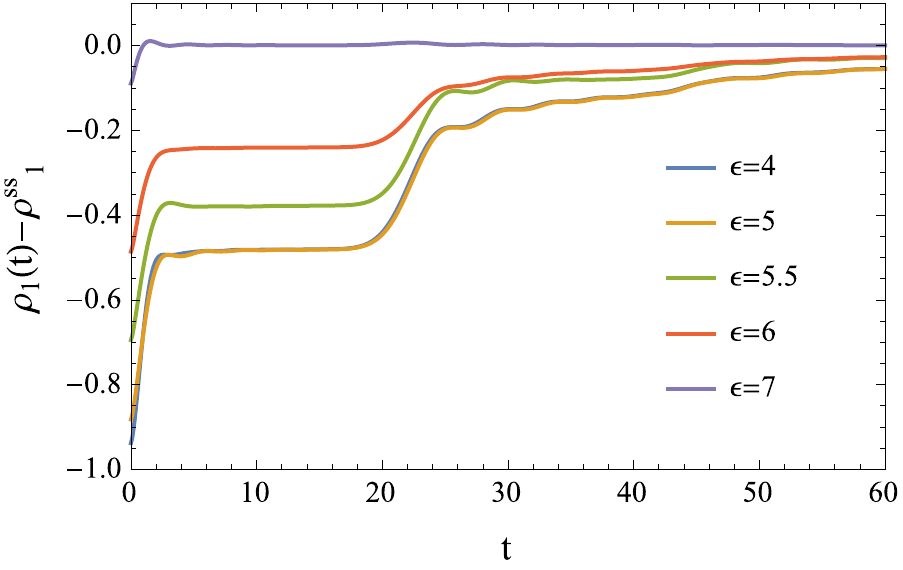}
\end{center}
\caption{(Color Online) Behavior of the time-dependent electron density at site 1,
 $\rho_1(t)$, by time for several $\epsilon$. We set the parameters as $N=20,\
V_L=V_R=6,\ \Gamma_L=\Gamma_R=\gamma=0.5$. There are three regions in
 the case: $\epsilon > V + 2 \gamma = 7$, $\epsilon < V - 2 \gamma = 5$,
 and $5 < \epsilon < 7$.}
\label{fig:rho_e}
\end{minipage}
\hfil
\begin{minipage}[t]{0.45\hsize}
\begin{center}
 \includegraphics[width=0.95\hsize]{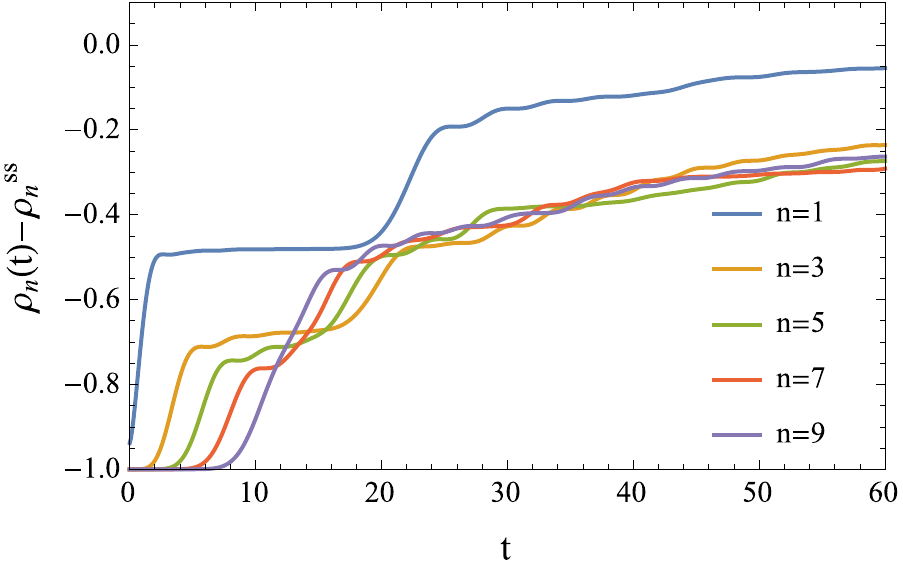}
\end{center}
\caption{(Color Online) The behavior of the time-dependent electron
 density $\rho_n(t)$ by time $t$ for several site indices $n$: $n=1,3,5,7,9$.}
\label{fig:rho_n}
\end{minipage}
\end{figure}

First we consider the case $0<l<1$, where there is no special eigenvalue. As we prove in Appendix C, for the case of $N \rightarrow \infty$, we can express the $n$th diagonal part of $\mathbf{\rho}^{(1,1)}(t)$ as
\begin{align}
 \mathbf{\rho}^{(1,1)}_n(t) 
&= \frac{i}{2 \pi} \frac{\gamma}{l} \int^{\infty}_{t} ds
 \frac{e^{i(\epsilon-V)s}}{s} 
 \int^{\infty}_s du e^{\gamma(l-\frac{1}{l})(u-s)} h_{nn}(u, l),
\label{eq:20}
\end{align}
where the integrand is defined as 
\begin{align*}
h_{nm}(s,l) &= I_{n-m}(-2\gamma is)-I_{n+m}(-2\gamma is) - l^2 (
I_{n-m}(-2\gamma is) - I_{n+m-2}(-2\gamma is)) \\
& \qquad + i 2l (I'_{n-m}(-2\gamma is) - I_{n+m-1}(-2\gamma is)).
\end{align*}
$I_n(s)$ is the
modified Bessel function of order $n$\cite{abramowitz1988handbook}. In Fig. \ref{fig:rho11}, we check the correctness of
the expression (\ref{eq:20}) by comparing the behavior of (\ref{eq:20}) with the numerical result
calculated from the definition (\ref{eq:19.1}). The figure shows that there
is good agreement between the analytical and the numerical result. 
In the similar way of deriving $\mathbf{\rho}^{(1,1)}(t)$, we can
calculate $\mathbf{\rho}^{(1,2)}(t)$ for large $N$, whose expression
changes according to the parameters $\epsilon$, $V$, and $\gamma$. For the case of $\epsilon > V + 2 \gamma$, which means that the on-site energy is large enough, $\mathbf{\rho}^{(1,2)}(t)$ is written as
\begin{equation}
 \mathbf{\rho}^{(1,2)}_n(t) = - \frac{i}{2 \pi} \sum^N_{m=1} A_{nm}(t) B_{mn}(t), 
\label{eq:21}
\end{equation}
where $A_{nm}(t)$ and $B_{mn}(t)$ are expressed as
\begin{align}
 A_{nm}(t) &= \sum_{k} [\mathbf{R}_k]_{nm}  e^{- i \lambda_k t} \notag \\
 &  \xrightarrow{N \rightarrow \infty}  \frac{\gamma}{l} e^{-i \epsilon t}
 \int^{\infty}_t ds e^{\gamma(l-\frac{1}{l})(s-t)} h_{nm}(s, l),
\label{eq:22}
\end{align}
\begin{align}
B_{mn}(t) &= \sum_{l} [\mathbf{R}^*_l]_{mn}  e^{ i \lambda^*_l t}
 E_1(it(\lambda^*_l - V)) \notag \\
 & \xrightarrow{ N \rightarrow \infty} \frac{\gamma}{l} e^{i \epsilon t} \int^{\infty}_t ds
 \frac{e^{-i(\epsilon- V)}s}{s} \int^s_{-\infty} du
 e^{-\gamma(l-\frac{1}{l})(u-s)} h^*_{mn}(u-t, l).
\label{eq:23}
\end{align}
A short derivation of the expressions (\ref{eq:22}) and (\ref{eq:23})
is in Appendix C. Next we consider the case of $\epsilon < V + 2
\gamma$. In this situation, a phase term arises from $E_1(it(\lambda^*_l
- V))$ in the definition of $B_{mn}(t)$, (\ref{eq:23}). To understand this
fact, let us consider the distribution of the eigenvalues $ \lambda^*_k - V$, which we can obtain from Fig. \ref{fig:eigendis}, the distribution of eigenvalues $- \lambda_k$, by folding its real value
with respect to the imaginary axis and shifting $-V$ along the real
axis. In this case $\epsilon < V + 2
\gamma$, the real parts of several eigenvalues $\mathrm{Re}(\lambda^*_k - V) \eqsim \epsilon + 2 \gamma \cos k - V$ are negative. By considering the fact that multiplying the imaginary number $i$ means
rotating 90 degrees around the origin of the complex plane, the
eigenvalues which satisfy $\mathrm{Re}(\lambda^*_k - V) < 0$, cross the
branch cut of $E_1(i(\lambda^*_k - V))$ on the negative real line. This
leads to the appearance of the phase factor $2\pi i$ for several terms in
the summation of $l$ (\ref{eq:21}). In
the following, we consider the case $\epsilon < V - 2 \gamma$ for simple
calculation. In the case of $\epsilon < V - 2 \gamma$, all the
arguments in $E_1(it(\lambda^*_l
- V))$ cross the branch cut. Therefore, $\mathbf{\rho}^{(1,2)}_n(t)$ changes from (\ref{eq:22}) to
\begin{equation}
 \mathbf{\rho}^{(1,2)}_n(t) = - \frac{i}{2 \pi} \sum^N_{m=1}
  (A_{nm}(t) B_{mn}(t) - 2 \pi i |A_{nm}(t)|^2).
\label{eq:24}
\end{equation}
We discuss the behavior of $\mathbf{\rho}^{(1,2)}_n(t)$ for $
V- 2 \gamma <\epsilon < V + 2 \gamma$ later in Fig. \ref{fig:rho_e}. The second term in (\ref{eq:24}), $\sum_m |A_{nm}(t)|^2$,
causes two-step decay. First we check this fact by numerically
calculating the definition of $\rho^{(1)}(t)$ and analytically
understand it later. We compute $\rho^{(1)}(t)$ instead of $\rho^{(1,2)}(t)$ because $\rho^{(1,1)}(t)$
converges to 0 very fast as Fig. {\ref{fig:rho11}} shows. The result is Fig. \ref{fig:rho1}. We numerically calculate
$\rho^{(1)}_n(t)$ from the
definition (\ref{eq:19}) by diagonalizing the non-hermitian matrix for
several combinations of the left coupling strength $0 \leq l < 1$ and site
index $n$. We set the parameters as $N=20,\  V_L=V_R=6,\ \epsilon=3,\
\Gamma_R=\gamma=0.5$. From the graph, we can see the following. For the case $l=0$,
where there is no coupling between the left reservoir and site 1,
$\rho^{(1)}_n(t)$ has two-step decay and the start time of each decay
depends on the site index $n$. For example, first decay occurs around the time $t=20+1-5=16$ and the
second starts around the time $t=20+1+5=26$ for the case $n=5$. In
contrast, for the case $n=15$, first decay occurs about the time
$t=20+1-15=6$ and the second occurs about the time
$t=20+1+15=36$. Therefore, we can expect that the start times change with the site index
$n$. For the case of $l \neq 0$, however, the start times of the
two-step decay are the same, $t=5$ and $t=15+1=16$, for both cases $n=5$ and
$n=15$. In the following, we analytically 
understand these facts above. First we consider the case $l=0$, where we can simply express $|A_{nm}(t)|^2$ as
\begin{align}
 |A_{nm}(t)|^2 &= |h_{nm}(t,0)|^2 \notag \\
 &= ((-1)^n J_{m-n}( 2 \gamma t) - J_{m+n}( 2 \gamma t))^2. 
\label{eq:24.1}
\end{align}
$J_n(t)$ is the Bessel function of the first kind. To obtain the
expression in the second line, we use the relation $I_n(t)= i^{-n}
J_n(it)$. We derive (\ref{eq:24.1}) from the definition in the same way as
(\ref{eq:22}). We can obtain the same result by taking the limit $l \rightarrow
0$ of (\ref{eq:22}) and replacing $\lim_{l \rightarrow 0}
e^{-\frac{t}{r}} /l$ for $t>0$ as $\delta(t)$. This replacement is not
mathematically true because $\int^{\infty}_{-\infty} dt \lim_{r \rightarrow 0}
e^{-\frac{t}{l}} /l \neq 1$, however, we can obtain the same expression in a simpler way. With the relation about the product of Bessel function
$\sum^{\infty}_{-\infty} J_n J_{n-k} = \delta_{0,k}$, we can calculate the
sum of (\ref{eq:24.1}) as
\begin{align}
 \sum^N_{m=1} |A_{nm}(t)|^2 &= \sum^{\infty}_{m=1} |A_{nm}(t)|^2 -
 \sum^{\infty}_{m=N+1} |A_{nm}(t)|^2 \notag \\
 &= 1 - \sum^{\infty}_{m=N+1} |A_{nm}(t)|^2 \notag \\
&= 1 - \sum^{\infty}_{m=N+1} \left( J^2_{m-n}( 2 \gamma t) + J^2_{m+n}( 2 \gamma t) - 2 (-1)^n
 J_{m-n}( 2 \gamma t) J_{m+n}( 2 \gamma t) \right). 
\label{eq:24.3}
\end{align}
From this expression, we can see that the sum $\sum^N_{n=1}
|A_{nm}(t)|^2$ has two-step decay. To understand this, we use the fact that the products of
Bessel function $J^2_n(t)$ and $J_{m-n}(t) J_{m+n}(t)$ take values which are almost 0 until $t
\eqsim n,\ m-n$ respectively. We prove this in Appendix C. We consider
the lowest order of each term in (\ref{eq:24.3}): $J^2_{N+1-n}(2\gamma t)$, $J^2_{N+1+n}(2\gamma t)$, and $J_{N+1-n}(2\gamma t) J_{N+1+n}(2\gamma t)$. Before the time $t < t_1 =  \frac{N+1-n}{2\gamma}$, all of the three terms take
0. The first and the third term have non-zero value in $ t_1 \leq t < t_2 =
\frac{N+1+n}{2\gamma}$. Around the time $t=t_2$, the second term start to have
non-zero value and all the terms have finite values after the time. From
the discussion above, we can conclude that the two decays start at
$t_1=\frac{N+1-n}{2\gamma}$ and $t_2=\frac{N+1+n}{2\gamma}$. This is in good agreement with the numerical
results of $\gamma=0.5$ in Fig. \ref{fig:rho1}. We can explain the fact above in more
intuitive way using Fig. \ref{fig:dg_transport}. 
We focus on the site $n$. Since the transport
is ballistic, the time which a electron need to move $i$ sites linearly
grows with the number $i$\cite{antal1999transport} with the
normalization of $2 \gamma$. Therefore, it takes time $t_1$ for an
electron from the right reservoir in Fig. \ref{fig:dg_transport} $(a)$
to reach site $n$ and the electron density at site $n$ changes at the
time $t_1$. After that, the
electron is reflected at the left most dot around $t=N$. Then the electron density
at site $n$ changes again at the time $t_2$ when the reflected electron comes
back to the site $n$. This is the reason why the electron density
changes at the times $t_1$ and $t_2$. With this understanding, we can study the case
of $l>0$, Fig. \ref{fig:dg_transport} $(b)$. In this case, there are couplings between the reservoirs and
both edges of the sites. Therefore, electrons are transported from the
both reservoirs. At the time $t=\frac{n}{2\gamma}$ and
$t=\frac{N-n+1}{2\gamma}$, the electrons from the reservoirs arrive at the
site $n$ and the quantity changes $\rho^{(1)}_n(t)$. This is also
compatible with the facts obtained from Fig. \ref{fig:rho1}.

By carrying out a similar calculation as $\mathbf{\rho}^{(1)}(t)$, we
can obtain an expression of $\mathbf{\rho}^{(2)}(t)$. From the expression, we can see
that the
behavior of $\mathbf{\rho}^{(2)}(t)$ is almost the same as
$\mathbf{\rho}^{(1, 2)}(t)$. Similar to the analysis of
$\mathbf{\rho}^{(1)}(t)$, we consider a symmetric bias voltage
$V_L=V_R=V$ and a large bias case
$V \gg 4 \gamma$ for simple calculation. In this case, we have the expression as 
\begin{align}
 \mathbf{\rho}^{(2)}(t) &= \sum_{\alpha=\{L,R \}}
 \mathbf{\rho}^{(2)}_{\alpha}(t) \notag \\
 &= V^2 \int^{\infty}_{-\infty} \frac{d \omega}{2 \pi} f(\omega-\mu)
 \sum_{k,l} \mathbf{R}_k \mathbf{\Gamma} \mathbf{R}^*_l e^{- i
 (\lambda_k - \lambda^*_l) t}
 \frac{1}{(\omega-\lambda_k)(\omega+V-\lambda_k)(\omega+V-\lambda^*_l)(\omega-\lambda^*_l)} \notag \\
&\eqsim \frac{1}{2\pi} \sum_{k,l} \mathbf{R}_k   \mathbf{\Gamma} \mathbf{R}^*_l e^{- i
 (\lambda_k - \lambda^*_l)t} \frac{1}{\lambda_k - \lambda^*_l}(F_1(\lambda_k) - F_1(\lambda^*_l) + F_1(\lambda_k - V) - F_1(\lambda^*_l - V)).
\label{eq:25}
\end{align}
Details of the derivation are in Appendix C. We use the function
$F_1(\lambda)$ defined in steady case again. For
$\beta=\infty$, we can calculate $F_1(\lambda)$ of Eq. (\ref{eq:25}) in the same way as (\ref{eq:14.1.1}). The result is 
\begin{align}
 \mathbf{\rho}^{(2)}(t) 
&= \frac{1}{2\pi} \sum_{k,l} \mathbf{R}_k  \mathbf{\Gamma} \mathbf{R}^*_l e^{- i
 (\lambda_k - \lambda^*_l)t} \times \notag \\
&\qquad \frac{1}{\lambda_k - \lambda^*_l} (\log(-\lambda_k) - \log(-\lambda^*_l) +  \log(V-\lambda_k) - \log(V-\lambda^*_l) - 4 \pi i),
\label{eq:26}
\end{align}
For the case of large $N$ and $\epsilon > 2 \gamma$, we have the relation $\log(-\lambda_k) -
\log(-\lambda^*_l) \eqsim 2 \pi i$, which we can understand graphically from 
Fig. \ref{fig:eigendis}. Again $\eqsim$ means that we ignore the order
of $O(1/N)$. In a similar way, the following relation holds 
\begin{equation}
\log(V-\lambda_k) - \log(V-\lambda^*_l) \eqsim 
 \begin{cases}
  0 & \epsilon < V - 2\gamma, \notag \\
 2 \pi i & \epsilon > V + 2\gamma.
 \end{cases}
\label{eq:26.1}
\end{equation}
By substituting these relations into (\ref{eq:26}) and using $ -i \sum_{k,l} \mathbf{R}_k  \mathbf{\Gamma} 
 \mathbf{R}^*_l / (\lambda_k - \lambda^*_l) e^{- i
 (\lambda_k - \lambda^*_l)t} =  \sum_{k,l} \mathbf{R}_k  \mathbf{R}^*_l / (\lambda_k - \lambda^*_l) e^{- i
 (\lambda_k - \lambda^*_l)t}$, which we can prove in a similar way as (\ref{eq:B1}) in Appendix B, we have
\begin{equation}
 \mathbf{\rho}^{(2)}_n(t) =
 \begin{cases}
 \sum^N_{m=1} |A_{nm}(t)|^2 & \epsilon < V - 2\gamma,  \\
 0  & \epsilon > V + 2\gamma.
 \end{cases}
\label{eq:rho2}
\end{equation}
Details of the derivation is in Appendix C. Again, the same quantity $\sum^N_{m=1} |A_{nm}(t)|^2$ as in (\ref{eq:24})
appears. Therefore, the two-step decay also appears in
$\mathbf{\rho}^{(2)}_n(t)$ in the similar way as
$\mathbf{\rho}^{(1)}_n(t)$. By substituting (\ref{eq:20}), (\ref{eq:21}), and
(\ref{eq:26.1}) into (\ref{eq:16}) with (\ref{eq:19}), we can obtain an analytical expression of the
time-dependent part of the electron density. Based on the discussion above, we
investigate the time-dependent part of the electron density itself for the case where there is no
special eigenvalue with numerical calculation in Fig. {\ref{fig:rho_e}} and
Fig. \ref{fig:rho_n}. Below, we simply call
the time-dependent part of the electron density as the time-dependent
electron density because the effect of the steady-state value is just
constant shifting. We confirm that the results obtained for the part of
the electron density $\rho^{(1,2)}(t)$ also hold for the electron density. 
In Fig. \ref{fig:rho_e}, we numerically compute the
time-dependent electron density at site 1 from the
definition (\ref{eq:16}) for several $\epsilon$ and see the dependence
of the two-step decay on $\epsilon$. We set the parameters as $N=20,\
V_L=V_R=6,\ \Gamma_L=\Gamma_R=\gamma=0.5$. As we expect, the electron
density has two-step decay for $\epsilon < V + 2 \gamma$. For the case
of $ \epsilon < V - 2 \gamma$, the transient behaviors of the electron
density for several $\epsilon$ are the same. We can understand
this fact from the expressions (\ref{eq:22}) and (\ref{eq:24}). Because
the term of on-site energy $\epsilon$ appears in (\ref{eq:22}) as a phase factor, the
effect disappears in (\ref{eq:24}) by taking the norm of $A_{nm}(t)$ and therefore the behavior of the time-dependent electron density for $ \epsilon < V - 2 \gamma$ is the same. In
the case $ V - 2 \gamma < \epsilon < V + 2 \gamma$, however, the
behavior of the electron density changes with the on-site energy. As the
energy increases to $\epsilon = V + 2 \gamma$, the value of the
electron density decays. This is because the number of the terms
$B_{nm}$ in (\ref{eq:21}) with the phase factor $2 \pi i$ decreases. In the
case of $ \epsilon > V + 2 \gamma$, the electron density quickly
converges towards 0. This behavior matches with our expectation obtained
in the discussion of $\sum_n |A_{nm}|^2(t)$. We also see the behavior of
the time-dependent electron density for several site $n$ in
Fig. \ref{fig:rho_n}. From this figure, we can confirm that the times
when the two-step decays occur change with site index according to $t_1 = \frac{N+1- n}{2\gamma}$ and
$t_2 = \frac{N+1+n}{2 \gamma}$, which are our theoretical prediction from (\ref{eq:24.3}).

Until now we only consider the case where there is no special
eigenvalue. We study the time-dependent electron density including the
effect of the special eigenvalue. Because
of the special eigenvalue, a new term appears in the expression for the electron density and it changes the dependence of the time-dependent
electron density on the boundary parameters. In the following analytical
study, we only focus on the behavior of the $A_{nm}$ because it
mainly determines the behavior of the time-dependent electron density as
we see it in the discussion of the case $r,l<1$ above. For simple analysis, we
consider the situation of $l=0, \ r>1$, where the special eigenvalue $\Lambda_r$ exists. For large
$N$, we can separate the relation (\ref{eq:10.4}) into
\begin{equation}
 \sum_{k} \mathbf{R}_k = \sum_{k:\ \mathrm{normal}} \mathbf{R}_k +
  \mathbf{R}_{\Lambda_r}.
\label{eq:27}
\end{equation}
By substituting (\ref{eq:27}) into the definition of $A_{nm}$
(\ref{eq:22}), we obtain the
expression for $A_{nm}(t)$ for $l=0,\ r>1$ as
\begin{align}
 A_{nm} &= \sum_{k:\ \mathrm{normal}} [\mathbf{R}_k]_{nm}  e^{- i
 \lambda_k t} + [\mathbf{R}_{\Lambda_r}]_{nm} e^{- i
 \Lambda_l t} \notag \\
 &\eqsim  e^{-i \epsilon t} \left( h_{nm}(t,0) - (1+r^2)
 (-ir)^{n+m-2(N+1)} e^{-\gamma(r-\frac{1}{r}) t }
 \right),
\label{eq:28}
\end{align}
where we only focus on the largest order of $r$. The detail of the derivation is in Appendix C. By comparing the
 expression (\ref{eq:28}) with the expression (\ref{eq:24.1}) of
 $A_{nm}$ for $l=0$ and $r<1$, we can see that an additional term $-
 (1+r^2) (-ir)^{n+m-2(N+1)} e^{-\gamma(r-\frac{1}{r}) t}$ appears due to the special
 eigenvalue $\Lambda_r$. When $n$ and $m$ are order of $O(N)$, this term
 remains in the limit $N \rightarrow \infty$. Because the common term $h_{nm}(t,0)$ does not
 depend on the boundary coupling parameter $r$, the additional term
 changes dependence on the
 boundary parameter at $r=1$ and cause an interesting behavior as we see
 blow.

We can see effects of the additional term in (\ref{eq:28}) from direct numerical
computation of the time-dependent electron density using (\ref{eq:16}). To see the dependence on the
boundary parameter, we fix $t$ at a time. At the time, we change the boundary
parameter $\Gamma_R$ and see the dependence of the electron density on
the boundary parameter. The result is Fig.\ref{fig:phasetrans1}. The red
points in the figure express extreme values of the graphs obtained
numerically. 
From the
graphs, we can notice that the dependence on the boundary parameter, or
$\frac{d \rho_n(t)}{dr}$, changes at $\Gamma_R =
2\gamma$, which is compatible with our analytical expectation above. In the region where
$\Gamma_R<2\gamma$, $\frac{d \rho_n(t)}{dr}$ is positive. This means
that the convergence speed of the electron density to steady state 
becomes fast by increasing $\Gamma_R$ since we calculate the difference
from the electron density in the steady state. This is intuitively natural
result because the boundary parameter determines how easy particles can
flow into the sites. In contrast,
the derivative is negative for $\Gamma_R>2\gamma$, or the convergence speed
becomes slower from the point $r=1$ by increasing the value of the boundary
parameter. Unlike the case of $r<1$, this is contrary to our intuition. We can physically
understand this behavior as follows. When the boundary parameter is smaller than the coupling parameter between the dots $\Gamma_R < 2 \gamma$, particles can hop to the next site from the rightmost site more easily than the rightmost site from the right reservoir $R$. Therefore, a bottleneck appears between the right most site and the right reservoir and convergence speed becomes fast by increasing the $\Gamma_R$. In contrast, for the case $\Gamma_R > 2 \gamma$, particles are stuck not between the rightmost site and the right reservoir but between the sites. In this situation, by increasing the boundary parameter, outflow of particles from the rightmost site to the right reservoir is likely to happen. This means that the convergence speed becomes slower by changing the boundary parameter. In this way, we can understand the change of dependence of the time-dependent electron density on the boundary parameter. We note that the behavior for
$\Gamma_R<2\gamma$ seems not to match our theoretical prediction above, but this difference appears to be due to finite size effects. In the limit
$N\rightarrow \infty$, the time-dependent electron density does not
depend on boundary parameter for the case $r,l<1$ as in
(\ref{eq:28}). However, in the finite case, the effect of the boundary
parameter remains and the dependence exists. Finally we can check the change of dependence on the boundary parameters for the case where both boundary parameters $\Gamma_L$ and
$\Gamma_R$ exist as in Fig. \ref{fig:phasetrans2}.

In this way, the dependence of time-dependent electron density on the
boundary parameter changes at $r,l=1$, where the special
eigenvalue of the non-hermitian matrix (\ref{eq:8}) appears. This means
that we can properly manipulate speed of decay by changing boundary
parameter. Though we can simply consider that the convergence speed to steady
state becomes faster as we increase the boundary parameters, the fastest
decay actually happens at $\Gamma_L =
\Gamma_R = \gamma$. If the boundary parameters are larger than the
value, the decay speed becomes slow. This fact can be useful for quantum
computation, where coherence time is very important quantity.

\section{Conclusion}
We have investigated the $N$ tight-binding model coupled to
two reservoirs with the nonequilibrium Green function including initial
correlation to understand effects of the open boundaries. By carrying out concrete analytical investigation of the
formal expression (\ref{eq:2}) in our case, we obtained expressions for
the steady-state and the
time-dependent electron density for the limit $N \rightarrow \infty$ and $\beta = \infty$. We showed that the non-hermitian
matrix, which appeared in the expression of the electron density, could have
special eigenvalues whose imaginary parts did not vanish even in the
thermodynamic limit $N \rightarrow \infty$. From the simple
expression of the electron density in the steady state, we analytically
showed that the special
eigenvalues did not affect the qualitative behavior
of the electron density and the phase transition which has been observed
for the isolated tight-binding model also existed in our open case. In
contrast, we found that the open boundaries affect the qualitative
behavior of the time-dependent electron density. From the expression of
the time-dependent electron density, we saw that the electron
density has the two-step decay caused by open boundaries for the region
where the on-site energy is not large enough $\epsilon < V +
2\gamma$. Because of the boundary effect, particles flow and the time
lag between flow of particles from the left and the right reservoir
causes the two-step decay as in Fig. \ref{fig:dg_transport}. In
addition, we showed that the
dependence of convergent speed on the boundary couplings changed by the
special eigenvalues. Because the boundary parameters determine the
existence of the special eigenvalues, the qualitative change of the
convergent speed is a result of the open
boundaries. 

Until now, we have considered the regime where the Coulomb interaction
between electrons are irrelevant. Therefore, it is interesting to
investigate the transient dynamics of the model including the Coulomb
interaction and understand the effect of the interaction to the
dynamics. 

\section*{Acknowledgments}

The work of T. S. is supported by JSPS KAKENHI Grant Numbers JP16H06338, JP18H01141, JP18H03672, JP19K03665.


\appendix

\section{About Eigenvalue and Eigenvector of $\mathbf{h}^{eff}$}

\subsection{Derivation of the Characteristic Equation}
In this subsection, we derive the characteristic equation of
$\mathbf{h}^{eff}$ (\ref{eq:5}). The characteristic equation of $\mathbf{h}^{eff}$ (\ref{eq:5}) is written as 
\begin{equation*}
 \begin{pmatrix}
     \epsilon - \frac{i}{2} \Gamma_L & \gamma & 0 & \cdots & & & &\\
     \gamma & \epsilon & \gamma & 0 & \cdots & & & \\
     0 & \gamma & \epsilon & \gamma & \cdots & & & \\
     \vdots & & & & & & & \\
     0 & \cdots & & & & 0 & \gamma &  \epsilon - i \frac{i}{2} \Gamma_R \\
 \end{pmatrix}
 \begin{pmatrix}
     e_1 \\
     e_2 \\
     \vdots \\
     \\
     e_N \\
 \end{pmatrix} 
 = \lambda \begin{pmatrix}
     e_1 \\
     e_2 \\
     \vdots \\
     \\
     e_N \\
 \end{pmatrix}. 
\end{equation*}
By introducing $e_0 = - i \frac{\Gamma_L}{2\gamma} e_1=-i l e_1$ and
$e_{N+1} = -i \frac{\Gamma_R}{2\gamma} e_N = -ir e_N$, the
characteristic equation is expressed as 
\begin{equation*}
 \begin{pmatrix}
     \gamma & \epsilon & \gamma & \cdots & & & &\\
     0 & \gamma & \epsilon & \gamma & \cdots & & & \\
     0 & 0 & \gamma & \epsilon & \cdots & & & \\
     \vdots & & & & & & & \\
     0 & \cdots & & & 0 & \gamma  & \epsilon &  \gamma \\
 \end{pmatrix}
 \begin{pmatrix}
     e_0 \\
     e_1 \\
     \vdots \\
     \\
     e_{N+1} \\
 \end{pmatrix} 
 = \lambda \begin{pmatrix}
     e_1 \\
     e_2 \\
     \vdots \\
     \\
     e_{N} \\
 \end{pmatrix},
\end{equation*}
or, in the form of the recurrence relation, 
\begin{equation}
 \gamma e_{n-1} + \epsilon e_n + \gamma e_{n+1} = \lambda e_n,
\label{eq:A1}
\tag{A.1}
\end{equation}
for $n=1,2 \cdots N$. To obtain the solution of the recurrence relation Eq. (\ref{eq:A1}), we
rewrite Eq. (\ref{eq:A1}) in a form of $e_{n+1} - \alpha e_n = \beta
( e_n - \alpha e_{n-1})$. By comparing the coefficients $e_{n+1} - \alpha e_n = \beta
( e_n - \alpha e_{n-1})$ with Eq. (\ref{eq:A1}), we obtain the following
relation
\begin{equation}
  \alpha + \beta = \frac{\lambda - \epsilon}{\gamma},\ \alpha \beta = 1.
\label{eq:A2}
\tag{A.2}
\end{equation}
Therefore, $\alpha$ and $\beta$ which satisfy the relation (\ref{eq:A2}) are
expressed as $\alpha,\beta = (\lambda - \epsilon)/2\gamma \pm
\sqrt{\left( (\lambda-\epsilon)/2\gamma \right)^2 - 1}$. Here we use the fact
$\beta \neq \alpha$, which we can prove. Using $\alpha$ and
$\beta$, we can obtain the following expression 
\begin{align*}
 e_{n+1} - \alpha e_n &= \beta^n (e_1 - \alpha e_0) \\
 &= (\beta^n + \beta ^{n-1} i  l) e_1,
\end{align*}
where we use the boundary condition $e_0 = -il e_1$ and $\alpha \beta
=1$ from the relation (\ref{eq:A2}). In the similar
manner, we can obtain the relation $e_{n+1} - \beta e_n = (\alpha^n  +
\alpha^{n-1} i \beta l)e_1$. By taking the difference of the two relations, we obtain 
\begin{equation*}
 (\beta - \alpha) e_n = \{ \beta^n - \alpha^n + il (\beta ^{n-1} -
  \alpha^{n-1} ) \} e_1
\label{eq:A3}
\tag{A.3}
\end{equation*}
Until now, we use only one of the boundary condition $e_0 = -il
e_1$. Therefore, Eq. \ref{eq:A3} have to satisfy the other boundary
condition, $e_{N+1} = -ir e_N$. From this condition, we obtain
\begin{equation*}
 \frac{e_{N+1}}{e_N} = \frac{\beta^{n+1} - \alpha^{n+1} + il (\beta^{n} -
  \alpha^{n} )  }{ \beta^n - \alpha^n + il (\beta ^{n-1} -
  \alpha^{n-1} ) } = -ir,
\end{equation*}
where we use $e_1\neq 0$. If $e_1 = 0$, then $e_n=0$ for
$n=1,2,\cdots N$ holds from the boundary condition $e_1 = -il e_0$ and
the recurrence relation (\ref{eq:A1}), which contradicts the definition
of eigenvector. The equation is expressed as 
\begin{equation}
 \beta^{n+1} - \alpha^{n+1} + i(l+r) (\beta^{n} -
  \alpha^{n} ) -rl (\beta ^{n-1} -
  \alpha^{n-1} ) = 0,
\label{eq:A4}
\tag{A.4}
\end{equation} 
is the same as Eq. (\ref{eq:6}). Eigenvalues are determined from
$\alpha$ and $\beta$ through the relation (\ref{eq:A2}). 

\subsection{Derivation of the Expression for $\mathbf{R}_k$}
In this subsection, we derive Eq. (\ref{eq:10.3.1}) and
Eq. (\ref{eq:10.3.2}). By definition of $\mathbf{R}_k$ in
Eq. (\ref{eq:10.2}), we have 
\begin{align}
 [\mathbf{R}_k]_{nm}
&= \left[ \frac{\mathbf{r}_k \mathbf{r}^t_k}{\mathbf{r}^t_k \cdot
  \mathbf{r}_k} \right]_{nm} \notag \\
&= \frac{(\sin n \Theta_k + i l \sin
  (n-1) \Theta_k)(\sin m \Theta_k + i l \sin
  (m-1) \Theta_k))}{\sum^N_{n=1} (\sin n \Theta_k + i l \sin
  (n-1) \Theta_k)^2}.
\label{eq:A5}
\tag{A.5}
\end{align}
We can calculate denominator in (\ref{eq:A5}) as follows.
\begin{align}
 \sum^N_{n=1} (\sin n \Theta_k + i l \sin
  (n-1) \Theta_k)^2 
&= \sum^N_{n=1}  \sin^2 n \Theta_k - l^2 \sin^2
  (n-1) \Theta_k + i 2 l \sin n \Theta_k \sin
  (n-1) \Theta_k \notag \\
&= \frac{1}{2}(1 - \cos 2 n \Theta_k) - \frac{l^2}{2}(1 - \cos (2 n-1)
 \Theta_k) \notag \\
& \quad + i 2 l \frac{-1}{2} ( \cos 2 (n-1) \Theta_k - \cos 2 \Theta_k ) \notag \\
&= \frac{N}{2}(1 - l^2 + i 2 l \cos \Theta_k) \notag \\
&\quad - \frac{1}{2}
 \sum^N_{n=1} ( \cos 2 n \Theta_k - l^2 \cos 2 (n-1)  \Theta_k + i 2 l \cos
 (2 n-1) \Theta_k ).
\label{eq:A6}
\tag{A.6}
\end{align}
Here we use the relations $\sum^N_{n=0} T_{2n}(z) = \frac{1}{2}(1+U_{2N}(z))$
and $\sum^N_{n=0} T_{2n+1}(z) = \frac{1}{2} U_{2N+1}(z)$. $T_{n}(\cos \theta)= \cos n \theta$ is
the Chebyshev polynomial of the first kind. By applying these relations
to Eq. (\ref{eq:A6}), we obtain the expression for the denominator of
Eq. (\ref{eq:A5})
\begin{align}
\sum_{n} (\sin n \Theta_k + i l \sin
  (n-1) \Theta_k)^2 &=  \frac{N}{2}(1 - l^2 + i 2 l \cos \Theta_k)
 \notag \\
& \quad - \frac{1}{4}
 ( -1 - l^2 + U_{2 N} ( \cos \Theta_k) - l^2 U_{2(N-1)} ( \cos \Theta_k) \notag \\
& \qquad + i 2 l U_{2 N-1} ( \cos \Theta_k) ),
\label{eq:A7}
\tag{A.7}
\end{align}
We can prove that the second term in Eq. (\ref{eq:A7}) is order of
$O(N^0)$. Therefore, for large $N$, we obtain Eq. (\ref{eq:10.3.1})
\begin{equation}
 [\mathbf{R}_k]_{nm} \eqsim \frac{2}{N} \frac{(\sin n \Theta_k + i l \sin
  (n-1) \Theta_k)(\sin m \Theta_k + i l \sin
  (m-1) \Theta_k))}{1 - l^2 + i 2 l \cos \Theta_k}.
\label{eq:A8}
\tag{A.8}
\end{equation}

Next we prove the expression of $\mathbf{r}^{(n)}_{\Lambda_l}$ and $\mathbf{R}_{\Lambda_l}$, or Eq. (\ref{eq:10.2.1}) and Eq. (\ref{eq:10.3.2}). To calculate Eq. (\ref{eq:10.2.1}), we need the expression of $\sin \Theta_k$ for the normalized special eigenvalue $\alpha = - i l$, which we represent as $\sin \Theta_{\Lambda_l}$. Here we use the fact that the expression for the characteristic equation (\ref{eq:10.1}) can
be obtained by substituting $\alpha = e^{i\Theta_k}$
into Eq. (\ref{eq:7}).  Therefore, $\sin n \Theta_{\Lambda_l}$ is expressed as 
\begin{align}
 \sin n \Theta_{\Lambda_l}
&= \frac{(-il)^n - (-il)^{-n} }{2i}.
\label{eq:A.9}
\tag{A.9}
\end{align}
By substituting the relation (\ref{eq:A.9}) into (\ref{eq:10.2}), we
obtain the expression of the eigenvector for the special eigenvalue, Eq. (\ref{eq:10.2.1}),
\begin{equation}
  \mathbf{r}^{(n)}_{\Lambda_l} = - \frac{(-il)^{-n+1}}{2}(l+\frac{1}{l}). \notag
\end{equation}
From the expression for Eq. (\ref{eq:10.2.1}), the denominator of  for large $N$ is calculated as
\begin{equation}
 \mathbf{r}^t_{\Lambda_l} \mathbf{r}_{\Lambda_l} \eqsim \frac{1}{4}(1+l^2).
\label{eq:A.10}
\tag{A.10}
\end{equation}
By substituting (\ref{eq:10.2.1}) and (\ref{eq:A.10}) into the definition of $\mathbf{R}_k$ in Eq. (\ref{eq:10.4}), we obtain
\begin{equation}
 [\mathbf{R}_{\Lambda_l}]_{nm} \eqsim - (1 + l^2) (- i l)^{-(n+m)},
\notag
\end{equation}
which is Eq. (\ref{eq:10.3.2}).

\section{Derivation of Eq. (\ref{eq:rhost}) }
In this section, we prove 
\begin{align}
 \rho^{ss}_{phase} &:= -i \sum_{k,l} (\lambda_k - \lambda^*_l)^{-1} \mathbf{R}_k  \mathbf{\Gamma} 
 \mathbf{R}^*_l \notag \\
&= \mathbf{1},
\label{eq:B1}
\tag{B.1}
\end{align} 
and Eq. (\ref{eq:rhost}) in Section 3. We can prove Eq. (\ref{eq:B1}) as follows.
\begin{align}
 \rho^{ss}_{phase} &= -i \sum_{k,l} (\lambda_k - \lambda^*_l)^{-1}
 \mathbf{R}_k  \mathbf{\Gamma}  \mathbf{R}^*_l \notag \\
&= -i \sum_{l} (\mathbf{h}^{eff}- \lambda^*_l \mathbf{1})^{-1}  \mathbf{\Gamma} 
 \mathbf{R}^*_l \notag \\
&= 2 \sum_{l} (\mathbf{h}^{eff}- \lambda^*_l \mathbf{1})^{-1}  (\mathbf{h}^{eff} -
 \mathbf{h}) \mathbf{R}^*_l, \notag 
\end{align}
where we use $\sum_{k} \mathbf{R}_k =1$ from the first line to the
second line. From the second line to the third line, we apply the
relation $\mathbf{h}^{eff} = \mathbf{h} - \frac{i}{2}
\mathbf{\Gamma}$. The expression is written as
\begin{align}
 \rho^{ss}_{phase}
&= 2 \sum_{l} (\mathbf{h}^{eff}- \lambda^*_l \mathbf{1})^{-1}  ( (\mathbf{h}^{eff} - \lambda^*_l \mathbf{1}) +
 (\lambda^*_l \mathbf{1} - \mathbf{h}) ) \mathbf{R}^*_l \notag \\
&= 2 \sum_{l} \{ \mathbf{1} + (\mathbf{h}^{eff}- \lambda^*_l \mathbf{1})^{-1}
 (\lambda^*_l \mathbf{1} - \mathbf{h}) \} \mathbf{R}^*_l \notag \\
&= 2 \mathbf{1} + 2 \sum_{l} (\mathbf{h}^{eff}- \lambda^*_l \mathbf{1})^{-1}
 ((\mathbf{h}^{eff})^* - \mathbf{h})  \mathbf{R}^*_l \notag \\
&= 2 \mathbf{1} + i \sum_{l} (\mathbf{h}^{eff}- \lambda^*_l \mathbf{1})^{-1}
  \mathbf{\Gamma} \mathbf{R}^*_l \notag \\
&= 2 \mathbf{1} - \rho^{ss}_{phase}, 
\label{eq:B2}
\tag{B.2}
\end{align}
In the first line, we insert the identity $-
\lambda^*_l \mathbf{1} + \lambda^*_l \mathbf{1}=0$. From the fourth line
to final line, we use the definition of $\rho^{ss}_{phase}$ in
Eq. (\ref{eq:B1}). Eq. (\ref{eq:B2}) shows that Eq. (\ref{eq:B1})
holds. In a similar way, we can prove Eq. (\ref{eq:rhost}). We start
from Eq. (\ref{eq:14.1}),
\begin{align}
\mathbf{\rho}^{ss}
&= 1 + \frac{1}{2 \pi} \sum_{\alpha} \sum_{k,l}  \left[  \mathbf{R}_k  \mathbf{\Gamma}_{\alpha} \left(
 \mathbf{R}_l \right)^* \right] \frac{1}{\lambda_k - \lambda^*_l}
 (\log( V_{\alpha}-\lambda_k) - \log( V_{\alpha} -\lambda^*_l))
 \notag \\
&=: 1 + \sum_{\alpha} \mathbf{\rho}^{ss}_{\alpha,1} +
 \mathbf{\rho}^{ss}_{\alpha,2 }.
\label{eq:B3}
\tag{B.3}
\end{align}
We can rewrite $\mathbf{\rho}^{ss}_{L, 1}$ as 
\begin{align}
 \mathbf{\rho}^{ss}_{L,1} 
&= \frac{1}{2 \pi} \sum_{k,l}   \mathbf{R}_k  \mathbf{\Gamma}_{\alpha} 
 \mathbf{R}^*_l  \frac{1}{\lambda_k - \lambda^*_l}
  \log( V_L-\lambda_k) \notag \\
&=\frac{1}{2 \pi} \sum_{l} \log( V_L- \mathbf{h}^{eff})
 (\mathbf{h}^{eff} - \lambda^*_l \mathbf{1})^{-1} \mathbf{\Gamma}_{L} \mathbf{R}^*_l \notag \\
&=\frac{1}{2 \pi} \sum_{l} \log( V_L- \mathbf{h}^{eff})
 (\mathbf{h}^{eff} - \lambda^*_l \mathbf{1})^{-1} (\mathbf{\Gamma} -
 \mathbf{\Gamma}_{R}) \mathbf{R}^*_l, \notag 
\end{align}
where we use the definition $\mathbf{\Gamma} = \mathbf{\Gamma}_{L} +
\mathbf{\Gamma}_{R}$. With the definition $\mathbf{h}^{eff} =
\mathbf{h} - \frac{i}{2} \mathbf{\Gamma}$ , we have 
\begin{align}
\mathbf{\rho}^{ss}_{L,1} 
&=\frac{1}{2 \pi} \sum_{l} \log( V_L- \mathbf{h}^{eff})
 (\mathbf{h}^{eff} - \lambda^*_l \mathbf{1})^{-1} \{2i (\mathbf{h}^{eff}
 - \lambda^*_l \mathbf{1} + \lambda^*_l \mathbf{1}
 - \mathbf{h} + \frac{i}{2} \mathbf{\Gamma}_{R})\} \mathbf{R}^*_l \notag \\
&= \frac{i}{\pi} \sum_{l} \log( V_L- \mathbf{h}^{eff})
 \{ \mathbf{1} + (\mathbf{h}^{eff} - \lambda^*_l \mathbf{1})^{-1} (
 \lambda^*_l \mathbf{1} -
 \mathbf{h} + \frac{i}{2} \mathbf{\Gamma}_{R}) \} \mathbf{R}^*_l
 \notag \\
&= \frac{i}{\pi} \sum_{l} \log( V_L- \mathbf{h}^{eff})
 \{ \mathbf{1} + (\mathbf{h}^{eff} - \lambda^*_l \mathbf{1})^{-1} (
  (\mathbf{h}^{eff})^* -
 \mathbf{h} + \frac{i}{2} \mathbf{\Gamma}_{R}) \} \mathbf{R}^*_l
 \notag \\
&= \frac{i}{\pi}  \log( V_L - \mathbf{h}^{eff}) -
 \mathbf{\rho}^{ss}_{L,1} -
 \frac{1}{\pi} \sum_{l} \log( V_L- \mathbf{h}^{eff})
  (\lambda_k \mathbf{1} - \lambda^*_l)^{-1} \mathbf{\Gamma}_{R} \mathbf{R}^*_l.
 \notag 
\end{align}
We insert $-
\lambda^*_l \mathbf{1} + \lambda^*_l \mathbf{1}=0$ in the first
line. From the third line to the final line, we use $(\mathbf{h}^{eff})^* -
 \mathbf{h} + \frac{i}{2} \mathbf{\Gamma}_{R} = \frac{i}{2}
 \mathbf{\Gamma}_{L} + i \mathbf{\Gamma}_{R}$. Therefore, we obtain
\begin{equation}
 \mathbf{\rho}^{ss}_{L,1} = \frac{i}{2 \pi}  \log( V_L -
  \mathbf{h}^{eff}) - \frac{1}{2 \pi} \sum_{l} \log( V_L- \mathbf{h}^{eff})
  (\lambda_k \mathbf{1} - \lambda^*_l)^{-1} \mathbf{\Gamma}_{R}
  \mathbf{R}^*_l
\label{eq:B4}
\tag{B.4}
\end{equation}
By carrying out the same calculation for $\mathbf{\rho}^{ss}_{R,1}$,
$\mathbf{\rho}^{ss}_{1}$ is expressed as
\begin{align}
 \mathbf{\rho}^{ss}_{1} 
&= \sum_{\alpha} \mathbf{\rho}^{ss}_{\alpha,1} \notag \\
&= \frac{i}{2 \pi}  \sum_{\alpha} \log( V_{\alpha} -
  \mathbf{h}^{eff}) - \frac{1}{2 \pi} \sum_{k,l} \mathbf{R}_k
  \frac{1}{\lambda_k  - \lambda^*_l} ( \log( V_L -
  \lambda_k) \mathbf{\Gamma}_{R} + \log( V_R -
  \lambda_k) \mathbf{\Gamma}_{L}) \mathbf{R}^*_l.
\label{eq:B5}
\tag{B.5}
\end{align}
In a similar way, we can obtain an expression of $\mathbf{\rho}^{ss}_{2}$
\begin{align}
 \mathbf{\rho}^{ss}_{2} 
&= - \frac{i}{2 \pi}  \sum_{\alpha} \log( V_{\alpha} -
  (\mathbf{h}^{eff})^* ) + \frac{1}{2 \pi} \sum_{k,l} \mathbf{R}_k
  \frac{1}{\lambda_k  - \lambda^*_l} ( \log( V_L -
  \lambda^*_l) \mathbf{\Gamma}_{R} + \log( V_R -
  \lambda^*_l) \mathbf{\Gamma}_{L}) \mathbf{R}^*_l.
\label{eq:B6}
\tag{B.6}
\end{align}
By substituting (\ref{eq:B5}) and (\ref{eq:B6}) for (\ref{eq:B3}), we have the expression for
the electron density in the steady state, (\ref{eq:rhost})
\begin{align}
\mathbf{\rho}^{ss} &= 1 + \frac{i}{2\pi} \sum_{\alpha} (
 \log( V_{\alpha} - \mathbf{h}^{eff} ) - \log( V_{\alpha} - (\mathbf{h}^{eff})^* )) \notag \\
& \quad - \frac{1}{2 \pi} \sum_{k,l} \mathbf{R}_k \frac{1}{\lambda_k
 - \lambda^*_l} \bigl( ( \log(V_L - \lambda_k ) - \log ( V_L - \lambda^*_l) )
 \mathbf{\Gamma}_{R} \notag \\
& \qquad +  ( \log(V_R - \lambda_k ) - \log ( V_R - \lambda^*_l))
 \mathbf{\Gamma}_{L}  \bigr)  \left( \mathbf{R}_l \right)^*. \notag
\end{align}

\section{About the derivation of the Analytical Expression for the
 Time-dependent Electron Density}

\subsection{Derivation of the Expression of $F_2(t,z)$ for $\beta= \infty$}
In this appendix, we first derive (\ref{eq:18}), the representation of $F_2(t,z)$ for $\beta= \infty$. By definition, we have the following expression
\begin{align}
 F_2(t,z) &= \int^{0}_{- \infty} \frac{d \omega}{2 \pi} \frac{e^{i \omega t}}{\omega - z} \notag \\
&= \int^{0}_{- \infty} \frac{d \omega}{2 \pi} \frac{\cos \omega}{\omega - tz} + i \int^{0}_{- \infty} \frac{d \omega}{2 \pi} \frac{\sin \omega}{\omega - tz} \notag \\
&= \frac{1}{2 \pi} (- g(tz) + i f(tz)),
\label{eq:C1}
\tag{C.1}
\end{align}
where we define $g(z)= \int^{\infty}_0 \frac{\cos \omega}{\omega + z}$ and $f(z)= \int^{\infty}_0 \frac{\sin \omega}{\omega + z}$. These functions are expressed with the first-order exponential integral with its principal value as
\begin{equation}
 g(z) - i f(z) = e^{iz} E_1(iz) \notag
\end{equation}
Note that this expression only holds for $ - \pi < \mathrm{arg} z < \frac{\pi}{2}$. By substituting the relation into (\ref{eq:C1}), we obtain 
\begin{equation}
 F_2(t,z) = - \frac{e^{i t z}}{2 \pi} E_1(itz).
\label{eq:C2}
\tag{C.2}
\end{equation}
for $ - \pi < \mathrm{arg} z < \frac{\pi}{2}$. For the case of $ \frac{\pi}{2} < \mathrm{arg} z < \pi$, or $\mathrm{Re} z <0$ and $\mathrm{Im} z >0$, $E_1(itz)$ changes to $E_1(itz) - 2 \pi i$. We understand this as follows. The exponential integral of first order with its principal value is expressed as 
\begin{equation*}
 E_1(z) = - \gamma - \mathrm{Log} z + \mathrm{Ein}(z),
\end{equation*}
where $\gamma$ is the Euler-Mascheroni constant and $\mathrm{Ein}(z)= \int^z_0 dt \frac{1-e^{-t}}{t}$ is holomorphic function. Therefore, the term $- 2 \pi i$ arises from $\log iz$ in the general value of $E_1(z)$ when $\mathrm{Re} z <0$ and $\mathrm{Im} z >0$. This is because $iz$ cross the branch of negative real line. Therefore, we obtain the expression (\ref{eq:18}).

\subsection{Derivation of the Expression for $\mathbf{\rho}^{(1,1)}_n(t)$, $\mathbf{\rho}^{(1,2)}_n(t)$, and $\mathbf{\rho}^{(2)}_n(t)$}

In this appendix, we derive (\ref{eq:20}), (\ref{eq:21}), (\ref{eq:25}) and
(\ref{eq:28}), which are the expression of
$\mathbf{\rho}^{(1,1)}_n(t)$, $\mathbf{\rho}^{(1,2)}_n(t)$, $\mathbf{\rho}^{(2)}_n(t)$ for $ 0 <
l < 1$, and $A_{nm}(t)$ for $l=0,\ r > 1$ respectively. First we show the derivation of $\mathbf{\rho}^{(1,1)}_n(t)$ for $l=0$ case, which enable us to
understand the derivation for $0<l<1$ easily. We start from the definition of
$\mathbf{\rho}^{(1,1)}_n(t)$ in (\ref{eq:19.1}). We use the expression of $[\mathbf{R_k}]_{nn}$ in
(\ref{eq:10.3.1}) and the eigenvalue $\lambda_k \eqsim
\epsilon + 2 \gamma \cos \Theta_k$ for large $N$. By substituting these expressions and (\ref{eq:18}) into the definition and taking $N \rightarrow \infty$, we have
\begin{align}
 \mathbf{\rho}^{(1,1)}_n(t) &= - i \sum_{k} [\mathbf{R}_k]_{nn} e^{- i (\lambda_k - V) t}  F_2(t, \lambda_k - V) \notag \\
 &= -i \sum_{k} \frac{2}{N}  \sin^2 n \theta e^{- i ( \epsilon + 2 \gamma \cos \Theta_k  - V) t}
 \times \notag \\
&\quad - \frac{e^{i t ( \epsilon + 2 \gamma \cos \Theta_k  - V)}}{2 \pi} E_1(it( \epsilon + 2 \gamma \cos \Theta_k  - V))  \notag \\
&\xrightarrow{N \rightarrow \infty}  \frac{i}{2 \pi} \frac{2}{\pi} \int^{\pi}_0 d \theta  \sin^2 n \theta  E_1 (i t (\epsilon + 2 \gamma \cos \theta -
 V)), 
\label{eq:C3}
\tag{C.3}
\end{align}
where we use $\Delta \Theta_k
= \Theta_k - \Theta_{k-1} = \frac{\pi}{N}$ for large $N$. 
To calculate the $\theta$-integral in (\ref{eq:C3}), we use the expression
of the exponential integral $E_1(itx)= \int^{\infty}_t ds
\frac{e^{- ixs}}{s}$. By substituting the expression into (\ref{eq:C3}),
we have the following expression 
\begin{align}
 \mathbf{\rho}^{(1,1)}_n(t) &= \frac{i}{2 \pi} \frac{2}{\pi} \int^{\pi}_0 d \theta  \sin^2 n \theta \int^{\infty}_t ds
\frac{e^{- i s (\epsilon + 2 \gamma \cos \theta -
 V)}}{s} \notag \\
&= \frac{i}{2 \pi} \frac{1}{\pi} \int^{\pi}_{- \pi} d \theta  \sin^2 n \theta \int^{\infty}_t ds
\frac{e^{- i s (\epsilon  - V )}}{s} e^{ - i s 2 \gamma \cos \theta} \notag \\
&= \frac{i}{2 \pi} \frac{1}{\pi} \oint_C \frac{dz}{iz}  \left(
 \frac{1}{2i}( z^n - z^{-n}) \right)^2 \int^{\infty}_t ds
\frac{e^{- i s (\epsilon  - V )}}{s} e^{ - i s 2 \gamma
 \frac{1}{2}(z+z^{-1})} \notag \\
&= \frac{i}{2 \pi} \frac{-1}{4 \pi i}  \int^{\infty}_t ds
\frac{e^{- i s (\epsilon  - V )}}{s} \oint_C \frac{dz}{z}  ( z^n -
 z^{-n} )^2 e^{ - i s \gamma (z+z^{-1})}, 
\label{eq:C4}
\tag{C.4}
\end{align}
where the contour $C$ is the unit circle. From the first line to the
second line, we use the fact that the integrand is even function. From
the second to the third line, we change variable from $\theta$ to
$z=e^{i \theta}$. We can calculate the complex integral in
(\ref{eq:C4}]) by using $e^{\frac{1}{2} s (z+\frac{1}{z})} =
\sum^{\infty}_{m = -\infty} I_m(s) z^m$, where $I_n(x)$ is the modified
Bessel function,  and the residue theorem. With this relation, we can
obtain the concrete expression of the complex integral as
\begin{align}
 \oint_C \frac{dz}{z} ( z^n - z^{-n})^2 e^{ - i s \gamma
 (z+\frac{1}{z})}
&= 2 \pi i \mathrm{Res}(z=0) \notag \\
&= 2 \pi i(I_{-2n}(-2i \gamma s) - 2I_0 (-2i \gamma s) + I_{2n}(-2i
 \gamma s)) \notag \\
&= 4 \pi i(I_{2n}(-2i \gamma s) - I_0 (-2i \gamma s)),
\label{eq:C5}
\tag{C.5}
\end{align}
where we use the relation $I_n(z) = I_{-n}(z)$ from the second line to
the third line. Using this result, we arrive at the expression of
$\mathbf{\rho}^{(1,1)}_n(t)$ for $r=0$
\begin{align}
 \mathbf{\rho}^{(1,1)}_n(t)
&= \frac{i}{2 \pi} \frac{-1}{4 \pi i}  \int^{ \infty}_t ds
\frac{e^{- i s (\epsilon  - V )}}{s} 2 \pi i \mathrm{Res}(z=0)
 \notag \\
&= \frac{i}{2 \pi} \frac{-1}{4 \pi i}  \int^{ \infty}_t ds
\frac{e^{- i s (\epsilon  - V )}}{s} 4 \pi i(I_{2n}(-2i \gamma s) - I_0 (-2i \gamma s)) \notag \\
&= \frac{i}{2 \pi} \ \int^{ \infty}_t ds
\frac{e^{- i s (\epsilon  - V )}}{s} (I_0 (-2i \gamma s) - I_{2n}(-2i \gamma s)). \notag
\end{align}
We calculate $\mathbf{\rho}^{(1,1)}_n(t)$ for
the case of $0 < l
<1$, (\ref{eq:20}). In the same way as deriving the first
line in (\ref{eq:C4}), we have the following expression for $0<l<1$
\begin{align}
\mathbf{\rho}^{(1,1)}_n(t)
&= \frac{i}{2 \pi} \frac{2}{\pi} \int^{\pi}_0 d \theta  \frac{(\sin n \theta + i l \sin
  (n-1) \theta)^2}{1-l^2 + 2 i l \cos \theta} E_1 (i t (\epsilon + 2 \gamma \cos \theta -
 V)) \notag \\
&= \frac{i}{2 \pi} \frac{2}{\pi} \int^{\pi}_0 d \theta  \frac{(\sin n \theta + i l \sin
  (n-1) \theta)^2}{1-l^2 + 2 i l \cos \theta} \int^{\infty}_t ds
\frac{e^{- i s (\epsilon + 2 \gamma \cos \theta -
 V)}}{s} \notag \\
&= \frac{i}{2 \pi} \frac{2}{\pi} \int^{\infty}_t ds
\frac{e^{- i s (\epsilon  - V )}}{s} \int^{\pi}_0 d \theta  \frac{(\sin n \theta + i l \sin
  (n-1) \theta)^2}{1-l^2 + 2 i l \cos \theta} e^{ - i s 2 \gamma \cos
 \theta}, 
\label{eq:C6}
\tag{C.6}
\end{align}
In contrast to (\ref{eq:C3}), the term $1-l^2 + 2 i l \cos \theta$
appears in the integration. To carry out $\theta$-integral in (\ref{eq:C6}), we express $e^{ - i s 2 \gamma \cos \theta}$ in (\ref{eq:C6}) as
\begin{align}
 \mathrm{exp} \left[ - i s 2 \gamma \cos \theta \right]
&= \mathrm{exp} \left[ - s \frac{\gamma}{l} \left( (1- l^2) - (1-l^2) + 2 il
 \cos \theta \right) \right]
 \notag \\
&= \mathrm{exp} \left[ s \frac{\gamma}{l}(1-l^2) \right] \mathrm{exp} \left[ - s \frac{\gamma}{l}( (1- l^2)
 + 2 il \cos \theta) \right] \notag \\
&= e^{s \frac{\gamma}{l}(1-l^2)} \int^{\infty}_s du \frac{\gamma}{l} (1-l^2 + 2 il \cos
 \theta ) e^{ - u \frac{\gamma}{l}( 1- l^2
 + 2 il \cos \theta)} \notag \\
&= \frac{\gamma}{l}  (1-l^2 + 2 il \cos \theta)  \int^{\infty}_s du e^{
 \gamma( l- \frac{1}{l}) (u-s)} e^{ - i u 2 \gamma \cos \theta},
\label{eq:C7}
\tag{C.7}
\end{align}
By substituting (\ref{eq:C7}) into (\ref{eq:C6}) and carrying out the
similar calculation as (\ref{eq:C5}), we obtain
\begin{align}
 \mathbf{\rho}^{(1,1)}_n(t) 
&= \frac{i}{2 \pi} \frac{2}{\pi} \int^{\infty}_t ds
\frac{e^{- i s (\epsilon  - V )}}{s} \int^{\pi}_0 d \theta  (\sin n
 \theta + i l \sin (n-1) \theta)^2 \times \notag \\
&\qquad \frac{\gamma}{l} \int^{\infty}_s du e^{
 \gamma( l- \frac{1}{l}) (u-s)} e^{ - i u 2 \gamma \cos \theta} \notag \\
&= \frac{i}{2 \pi} \frac{\gamma}{l} \int^{\infty}_t ds
\frac{e^{- i s (\epsilon  - V )}}{s} \int^{\infty}_s du e^{
 \gamma( l- \frac{1}{l}) (u-s)} \times \notag \\
&\qquad \frac{2}{\pi} \int^{\pi}_0 d \theta  (\sin n
 \theta + i l \sin (n-1) \theta)^2 e^{ - i u 2 \gamma \cos \theta}
 \notag \\
&= \frac{i}{2 \pi} \frac{\gamma}{l} \int^{\infty}_t ds
\frac{e^{- i s (\epsilon  - V )}}{s} \int^{\infty}_s du e^{
 \gamma( l- \frac{1}{l}) (u-s)} h_{nn}(u,l),
\label{eq:C8}
\tag{C.8}
\end{align}
which is (\ref{eq:20}). 
Next we calculate the expression for (\ref{eq:21}). For
the case $\epsilon < V + 2 \gamma$, by substituting the expression for
$F_2(z)$, (\ref{eq:18}), into the definition of
$\mathbf{\rho}^{(1,2)}(t)$, we obtain
\begin{align}
 \mathbf{\rho}^{(1,2)}_n(t) 
&=  i  \sum_{k,l} \sum_m [\mathbf{R}_k]_{nm} [\mathbf{R}^*_l]_{mn}
 e^{- i (\lambda_k - V)t} F_2(t, \lambda^*_l - V) \notag \\
&= - i  \sum_{k,l} \sum_m [\mathbf{R}_k]_{nm} [\mathbf{R}^*_l]_{mn}
 e^{- i (\lambda_k - V)t} \frac{e^{i t (\lambda^*_l - V)}}{2 \pi}
 E_1(it(\lambda^*_l - V)) \notag \\ 
&= - \frac{i}{ 2\pi} \sum_{k} [\mathbf{R}_k]_{nm} e^{- i \lambda_k t}
 \sum_{l} [\mathbf{R}^*_l]_{mn} e^{i \lambda^*_l t} E_1(it(\lambda^*_l -
 V)) \notag \\
&= - \frac{i}{2 \pi}\sum^N_{m=1} A_{nm}(t) B_{mn}(t), \notag
\end{align} 
where $A_{nm}(t)$ and $B_{mn}(t)$ are defined in (\ref{eq:22}) and
(\ref{eq:23}) respectively. By using the expression of
$[\mathbf{R_k}]_{nn}$ in (\ref{eq:10.3.1}) and the eigenvalue $\lambda_k
\eqsim \epsilon + 2 \gamma \cos \Theta_k$, we can obtain the following
representations of $A_{nm}(t)$ and $B_{mn}(t)$ for $0<l<1$ in $N
\rightarrow \infty$ as
\begin{gather}
 A_{nm}(t) = \frac{2}{\pi} \int^{\pi}_0 d \theta \frac{(\sin n \theta + i l \sin
  (n-1) \theta)(\sin m \theta + i l \sin
  (m-1) \theta)}{1-l^2 + 2 i l \cos \theta} e^{-i(\epsilon + 2\gamma
 \cos \theta) t}, \notag \\
B_{mn}(t) = \frac{2}{\pi} \int^{\pi}_0 d \theta \frac{(\sin m \theta - i l \sin
(m-1) \theta )( \sin n \theta - i l \sin (n-1) \theta) }{ 1-l^2 - 2 i l
 \cos \theta} e^{i(\epsilon + 2\gamma \cos \theta) t}
 E_1 (i t (\epsilon + 2 \gamma \cos \theta - V). \notag 
\end{gather}
By carrying out the calculation in the same way as (\ref{eq:C4}) and
(\ref{eq:C8}), we can obtain the expressions of $A_{nm}(t)$ and $B_{mn}(t)$ as
\begin{gather}
 A_{nm}(t) = \frac{\gamma}{l} e^{-i \epsilon t}
 \int^{\infty}_t ds e^{\gamma(l-\frac{1}{l})(s-t)} h_{nm}(s, l), \label{eq:C9}
\tag{C.9} \\
B_{mn}(t) = \frac{\gamma}{l} e^{i \epsilon t} \int^{\infty}_t ds
 \frac{e^{-is(\epsilon- V)}}{s} \int^s_{-\infty} du
 e^{-\gamma(l-\frac{1}{l})(u-s)} h^*_{mn}(u-t, l). \notag
\end{gather}
which are (\ref{eq:21}). 
In a similar way, we can derive the expression of $\mathbf{\rho}^{(2)}_n(t)$, (\ref{eq:rho2}) as shown in Section 5. In this appendix, we derive the expression of the third line in (\ref{eq:25}). We can express the fraction in the definition as 
\begin{align}
&\frac{V^2}{(\omega-\lambda_k)(\omega+V-\lambda_k)(\omega+V-\lambda^*_l)(\omega-\lambda^*_l)} \notag \\
&= \left( \frac{1}{\omega - \lambda_k} - \frac{1}{\omega + V - \lambda_k} \right) \left( \frac{1}{\omega - \lambda^*_l} - \frac{1}{\omega + V - \lambda^*_l} \right) \notag \\
&= \frac{1}{\lambda_k - \lambda^*_l} \left( \frac{1}{\omega - \lambda_k} - \frac{1}{\omega - \lambda^*_l} + \frac{1}{\omega +V - \lambda_k} - \frac{1}{\omega +V- \lambda^*_l} \right) \notag \\
& \qquad  - \frac{1}{\lambda_k - \lambda^*_l - V} \left(
 \frac{1}{\omega - \lambda_k} - \frac{1}{\omega +V-\lambda^*_l} \right)
 \notag \\
& \qquad
+ \frac{1}{\lambda_k - \lambda^*_l + V} \left(
 \frac{1}{\omega - \lambda^*_l} - \frac{1}{\omega +V-\lambda_k} \right)
\label{eq:C10}
\tag{C.10}
\end{align}
When $V \gg |\lambda_k - \lambda^*_l| \eqsim 4 \gamma$ holds and we ignore the term of $O(\frac{1}{V})$, the second and the third
terms vanish after the integration of $\omega$. Therefore, we can
reproduce the final line of (\ref{eq:25}).

Finally, we derive the expression for $A_{nm}(t)$ for $l=0, r>1$, where there is a
special eigenvalue $\Lambda_r$. For large $N$, we can calculate
$\mathbf{R}_{\Lambda_r}$ as 
\begin{align}
 [\mathbf{R}_{\Lambda_r}]_{nm}
&= \frac{\sin n \Theta_{\Lambda_r} \sin m \Theta_{\Lambda_r}}{\sum^N_{k=1}
 \sin^2 k \Theta_{\Lambda_r}} \notag \\
&= \frac{(-ir)^n - (-ir)^{-n}}{2i} \frac{(-ir)^m - (-ir)^{-m}}{2i}
 \frac{1}{\sum_k \left( \frac{(-ir)^k - (-ir)^{-k}}{2i} \right)^2}
 \notag \\
&\eqsim - (1+r^2) (-ir)^{n+m - 2(N+1)}.
\label{eq:C11}
\tag{C.11}
\end{align}
From the first line to the second
line, we use $\sin n \Theta_{\Lambda_r} = \frac{(-ir)^n -
(-ir)^{-n}}{2i}$, which we can obtain in the similar way of
(\ref{eq:A.9}). In the final line, we ignore terms of
$0(r^{-2N})$. Because $n$ and $m$ take $1,2 \cdots N$, the term
$(-ir)^{n+m - 2(N+1)}$ can remain for the large $N$ case.

\subsection{Behavior of the products of the Bessel functions}

In this subsection, we prove that the product $J_{m-n}(t) J_{m+n}(t) $ is almost 0 for $t < m-n$. We use this fact to understand the behavior of (\ref{eq:24.3}). From the definition of the Bessel function, we have the expression for the product of the Bessel function $J_{m-n}(t) J_{m+n}(t) $ as
\begin{equation}
 J_{m-n}(t) J_{m+n}(t) = \left( \frac{t}{2} \right)^{2m} \sum^{\infty}_{k=0} \frac{ (2 m + k + 1)_k \left( - \frac{1}{4} t^2\right)^2}{k! (m-n+k)! (m+n-k!)},
\label{eq:C12}
\tag{C.12}
\end{equation}
where $(\alpha)_k$ is Pochhammer's symbol: $(\alpha)_k = \alpha(\alpha+1)\cdots (\alpha+k-1)$. We can express (\ref{eq:C12}) as follows
\begin{align}
 J_{m-n}(t) J_{m+n}(t) &= \sum^{\infty}_{k=0} \frac{(-1)^k (2 m + k + 1)_k}{2^{2(m+k)} k!} \left( \frac{t}{m-n+k} \frac{t}{m-n+k-1} \cdots \frac{t}{1} \right) \notag \\
 & \qquad \left(  \frac{t}{m+n+k} \frac{t}{m+n+k-1} \cdots \frac{t}{m-n+k+1} \right),
\label{eq:C13}
\tag{C.13}
\end{align}
By dividing the denominators and numerators in the products in (\ref{eq:C13}) by $m-n$, we can see that the value of the products is almost 0 for $t< n-m$. In a similar manner, we can prove that the product $J^2_{m}(t)$ takes 0 for $t<n$.

\bibliographystyle{junsrt}

\end{document}